\documentclass{article}

\usepackage{amssymb}
\usepackage[figuresright]{rotating}
\usepackage{CJK,CJKspace}
\usepackage{indentfirst}
\usepackage{amsmath,amssymb,amsfonts,amsthm}
\usepackage{xcolor,graphicx,floatrow}
\usepackage{multicol}
\usepackage{hyperref}
\usepackage{color}
\usepackage{subfigure}
\usepackage{graphicx}
\usepackage{lscape}
\usepackage{epstopdf}
\usepackage{threeparttable}
\usepackage{amsmath}
\usepackage{dcolumn}
\usepackage{multirow}
\usepackage{booktabs}
\usepackage{listings}
\usepackage{longtable}
\usepackage{appendix}
\usepackage{geometry}
\hypersetup{linkcolor=blue,bookmarksopen=true}
\newtheorem{myDef}{Definition}[section]

\usepackage[footnotesize]{caption2}
\usepackage{booktabs,tabularx}
\title{Profitability of simple stationary technical trading rules with high-frequency data of Chinese Index Futures}

\author{Jing-Chao Chen\\School of Finance and Statistics\\ East China Normal University\\500 Dongchuan Road, Shanghai 200241, P.R. China
\and
Yu Zhou\\Antai College of Economics and Management\\ Shanghai Jiao Tong University\\800 Dongchuan Road, Shanghai 200241, P.R. China
\and
Xi Wang\\School of Finance and Statistics\\ East China Normal University\\500 Dongchuan Road, Shanghai 200241, P.R. China\\WangXi\_ECNU@126.com}

\begin{document}
\maketitle
\pagestyle{plain}

\begin{abstract}
Technical trading rules have been widely used by practitioners in financial markets for a long time. The profitability remains controversial and few consider the stationarity of technical indicators used in trading rules. We convert MA, KDJ and Bollinger bands into stationary processes and investigate the profitability of these trading rules by using 3 high-frequency data(15s,30s and 60s) of CSI300 Stock Index Futures from January 4th 2012 to December 31st 2016. Several performance and risk measures are adopted to assess the practical value of all trading rules directly while ADF-test is used to verify the stationarity and SPA test to check whether trading rules perform well due to intrinsic superiority or pure luck. The results show that there are several significant combinations of parameters for each indicator when transaction costs are not taken into consideration. Once transaction costs are included, trading profits will be eliminated completely. We also propose a method to reduce the risk of technical trading rules.


\end{abstract}

\section{Introduction}

Methods to analysis financial products generally can be divided into fundamental analysis and technical analysis. In contrary to fundamental analysis, technical analysis focuses on forecasting future price movement based on past market prices, turnover volume, and technical indicators. Some scholars criticize technical analysis since it violates the efficient market hypothesis. Jensen \cite{Jensen} proposes in his paper that " A market is efficient with respect to information set $\Theta_t$ if it is impossible to make economic profit by trading on the basis of information set $\Theta_t$". Some famous theories, such as random walk model and Black-Scholes pricing formula, are based on the market efficiency hypothesis. Some studies advocate this hypothesis and provide reports of negative evidence for technical analysis, including Levy \cite{Levy}, Malkiel \cite{Malkiel}, Bessembinder and Chan \cite{Chan}, and Olson \cite{Olson}.

As time goes by, however, there are increasing amount of evidences showing that the efficient market hypothesis is not valid in some cases. In academia, many literature advocate technical analysis has been published. Technical indicators are the most quantizable methods to conduct technical analysis, and have been widely used by financial practitioners to detect the market trends. Murphy \cite{Murphy} introduces formula and direction of many technical indicators in his book. H.Yu, et al. \cite{Yu} investigate moving average and trading range breakout rules on south Asian stock markets finding that trading rules have stronger predictive power in the emerging stock markets of Malaysia, Thailand, Indonesia, and the Philippines than in the more developed stock market of Singapore. Papailias and Thomakos \cite{Papailias} propose a moving average strategy with dynamic stop loss, and find it profitable for the price series of DJIA, S\&P500, and EUR/USD exchange rate. As for Chinese market, W.K.Wong, et al. \cite{Wong} apply the MA family (the Simple MA and its extensions, Exponential MA, Dual MA, Triple MA, MACD and TRIX) to Shanghai, Hong Kong and Taiwan markets and find that MA family achieves better performance than the buy-and-hold strategy regardless of transaction costs. Combining with exponential moving average, S.Chen, et al. \cite{Chen} study the predictive power of Japanese candlestick charting to in Chinese stock market and find that predictive power decreases as predicting period prolongs. H.L.Shi, et al. \cite{Shi1} investigate the profitability of loser, winner and contrarian portfolios in Chinese stock market, they discover that regardless the market is bullish or bearish, there exists significant long-term contrarian effect with holding horizons more than 12 months. Other work of H.L.Shi et al. about Chinese stock market can be found in Ref. \cite{Shi2,Shi3,Shi4}.L.X.Cui, et al. \cite{Cui} adopt DMD method to discover the
evolutionary patterns in stock market and apply it to Chinese A-share stock market finding that DMD algorithm can model the market patterns well in sideways market. J.J.Ma, et al. \cite{Ma} study the stock price fluctuation with intrinsic time perspective concluding that DC method can capture important fluctuations in Chinese stock market and gain profit due to the statistical property that average upturn overshoot size is bigger than average downturn directional change size.

In addition to literature mentioned above, there are still plenty of researches about technical indicators or trading rules. Few of them consider the stationarity of technical indicators. W.Liu, et al. \cite{Liu} transforms Bollinger bands to a stationary time series in their paper which may be the earliest study about stationary technical indicators. Based on similar method, more scholars started to construct different stationary technical indicators. Wang \& Zheng \cite{Z.D.Wang} find that based on the incremental stationary property of security price, several popular technical indicators can be proved stationary. They harness this property and discover several profitable high frequency trading strategies in China futures markets. Although the amount of literature about technical indicators is not little, there are few discuss the circumstance of high-frequency data within a minute. X.Wang, et al. \cite{X.Wang} propose a method to test the multi-dimensional stationary process and study the stock linkage in Chinese market with half second data. S.Bao, et al. \cite{Bao} study the application of stationary technical indicator in high-frequency trading based on MACD.

Besides investigating the efficiency of technical analysis, some scholars also focus on the reliability of strategies using technical indicators. Data snooping occurs when the same data is used more than once for the purpose of inference or model selection. This effect may lead to doubt of the reliability of strategies consisting of technical indicators. White \cite{White} illustrates what data-snooping is and creates the reality check(WRC) test to correct the data snooping effect. C.W.Chen, et al. \cite{C.W.Chen} test the same trading strategies as in Ref. \cite{Sullivan} in Asian stock markets and find that the WRC p-values of different markets were not the same. They also find that the predictive ability in Asian stock markets is not as good as that in the US. H.Zhu, et al. \cite{H.Zhu} investigate the profitability of moving average (MA) and trading range break (TRB) rules of Chinese stock exchange indexes and apply WRC to account for data snooping, finding that the best trading rule outperforms the buy-and-hold strategy when transaction costs are not taken into consideration. Hansen \cite{Hansen} proposes Superior Predictive Ability(SPA) test which is modified from WRC and more powerful under most circumstances. P.H.Hsu, et al. \cite{P.H.Hsu} further extends the WRC and SPA test into a stepwise SPA test and a stepwise Reality Check test. With these two extended tests, they examines the predictive ability of technical trading rules in emerging markets, and demonstrates that technical trading rules do have significant predictive abilities. S.Wang, et al. \cite{S.Wang} test the performance of technical trading rules in the Chinese markets based on SPA test and conclude that the predictive ability of technical trading rules appears when the market is less efficient.

In this paper, we will apply stationary technical indicators to Chinese stock index futures market and to examine whether any of
these technical trading rules would generate impressive profits by means of performance measure and risk measure. Adf-test will be used to verify the stationarity and SPA test to correct data snooping effect. The rest of this paper is arranged as follows. Section \ref{Section2} introduces the traditional and stationary technical trading rules as well as the data we use. The methodology of performance measure, risk measure and two statistical tests(Adf-test and SPA tes) are presented in Section \ref{Section3}. Section \ref{Section4} describes the empirical results and Section \ref{Section5} is the conclusion.

\section{Technical indicators\label{Section2} }

\subsection{Traditional technical indicators}

\subsubsection{Moving average\label{MA}}

Moving average is one of the most popular technical indicators, generating trading signals if the short-term moving average penetrates the long-term moving average. Generally speaking, moving average is a trend track indicator. When short-term MA is above(below) long-term MA, price will rise(fall) with higher probability. Denote an n-period moving average of price series $\{P_t\}$ by

$$MA(n,P)_t = \frac{P_t+P_{t-1}+\cdots+P_{t-n+1}}{n}  = \frac{\sum^t_{i=t-n+1}{P_i}}{n}.$$

If $n_l>n_s$, then $MA(n_s,P)_t$ presents the short-term moving average while $MA(n_l,P)_t$ presents the long-term moving average. More specifically, if $MA(n_s,P)_t$ rises above(fall below) $MA(n_l,P)_t$, a buy(sell) signal will be triggered. However, there are many ¡°fake¡± signals when short-term and long-term MA crosses each other frequently in a short period, which hardly bring any profit but increase the transaction costs. In order to avoid this situation, traders usually set up a filter zone with an up-band and a low-band, in which no trading signals will be generated. So the trading rules with MA and filter zone is defined as follows,

\begin{itemize}
\item Open long: $MA(n_s,P)_{t-1}-MA(n_l,P)_{t-1}\leq$up-band \& $MA(n_s,P)_{t}-MA(n_l,P)_{t}>$up-band;
\item Close long: $MA(n_s,P)_{t-1}-MA(n_l,P)_{t-1}>$up-band \& $MA(n_s,P)_{t}-MA(n_l,P)_{t}\leq$up-band;
\item Open short: $MA(n_s,P)_{t-1}-MA(n_l,P)_{t-1}>$low-band \& $MA(n_s,P)_{t}-MA(n_l,P)_{t}\leq$low-band;
\item Close short: $MA(n_s,P)_{t-1}-MA(n_l,P)_{t-1}\leq$low-band \& $MA(n_s,P)_{t}-MA(n_l,P)_{t}>$low-band.
\end{itemize}

\subsubsection{Stochastic oscillator(KDJ)\label{KDJ}}

George Lane promoted the stochastic oscillator indicator in the 1950s, which is a momentum indicator that uses support and resistance levels. The value of stochastic oscillator technical indicator is determined by the location of current price in relation to its price range over a period of time. The stochastic oscillator is displayed as two lines. The main line is called \%K. The second line, called \%D, is an iterative EMA of \%K. Sometimes traders also consider another line called \%J, which is a linear combination of \%K and \%D. Since the name of three lines, the stochastic oscillator indicator is also called KDJ. The algorithmic method is as follows. Denote $L_t$ and $H_t$ as the minimum and maximum price of the period of length $n$ respectively.

$$L_t = \min{ \{P_t, P_{t-1},\ldots, P_{t-n+1}\}},$$
$$H_t = \max{ \{P_t, P_{t-1},\ldots, P_{t-n+1}\}}.$$

Then we define $RSV_t$, $\%K_t$, $\%D_t$ and $\%J_t$ as follows where $m$, $k<n$. Usually we take $m$=3,$k$=3 and $n$=5, 9, or 14.

$$RSV_t = 100 \times \frac{P_t-L_t}{H_t-L_t},$$
$$\%K_t = \frac{m-1}{m+1}\%K_{t-1}+\frac{1}{m+1}RSV_{t},$$
$$\%D_t = \frac{k-1}{k+1}\%D_{t-1}+\frac{1}{k+1}K_{t},$$
$$\%J_t = 3\%K_t-2\%D_t.$$

There are several ways to interpret a stochastic oscillator. Three popular methods include:

\begin{itemize}
\item Buy when the oscillator (either \%K or \%D) falls below a specific level (e.g., 20) and then rises above that level. Sell when the oscillator rises above a specific level (e.g., 80) and then falls below that level;
\item Buy when the \%K line rises above the \%D line and sell when the \%K line falls below the \%D line;
\item Look for divergences. For instance: where prices are making a series of new highs and the stochastic oscillator is failing to surpass its previous highs.
\end{itemize}

\subsubsection{Bollinger bands\label{Boll}}

The Bollinger bands is a technical analysis tool invented by John Bollinger in the 1980s. An up-band and a low-band consist the Bollinger bands.

Denote an n-period exponential average(EMA) as follows.

$$EMA(n,P)_t = \frac{nP_t+(n-1)P_{t-1}+\cdots+P_{t-n+1}}{n+(n-1)+\cdots+1} = \frac{2}{n(n+1)}\sum^t_{i=t-n+1}{(n-t+i)P_i}.$$

Consider $EMA(n,P)_t$ to be the mid line of Bollinger bands, and denote the up-band and low-band as $EMA(n,P)_t+K\sigma_t$ and $EMA(n,P)_t-K\sigma_t$ respectively, where $K$ is a parameter and $\sigma$ is the standard deviation of the difference of price and $EMA(n,P)_t$.

$$\sigma(n,P)_t = \sqrt{\frac{1}{n}\sum^t_{i=t-n+1}{\left({P_i-EMA(n,P)_t}\right)^2}}.$$

When price is above the up-band, we consider the market is bull. On the contrary, when price is below the low-band, the market is bear. Therefore, buy signals will be generated when price rises above the up-band while sell signals will be generated when price falls below the low-band. In other word, no signal will be triggered when price is between the up-band and low-band.

\subsection{Stationary technical indicators}

\subsubsection{Brief introduction of stationary process\label{Stationary Process}}

\begin{myDef}
A stochastic process $\{X_t\}$ is called a stationary process, if for any $a>0$ and any finite time $t_1<t_2\ldots<t_n$, the random vector $\{X_{t_1}, X_{t_2}, \ldots, X_{t_n}\}$ has the same probability distribution as $\{X_{t_1+a}, X_{t_2+a}, \ldots, X_{t_n+a}\}$.
\label{Def1}
\end{myDef}

As Def.\ref{Def1} shows, the distribution of a stationary process will not change while time goes on. It is easy to see that, if $\{X_t\}$ is a stationary process and $f(x)$ is a function such that $\{f(X_t)\}$ is also a stochastic process, then $\{f(X_t)\}$ is a stationary process.

The definition of stationary process is perfect and too strict to satisfy. One can hardly verify a stochastic process is stationary. In actual applications, people care more about the mean and variation of a stochastic process. In this sense, weakly stationary process is more practical.

\begin{myDef}
A weakly stationary process satisfies the following two conditions: for each $c>0$,

\begin{itemize}
  \item $E[X_t]=E[X_{t+c}];$
  \item $E[X_s X_t]=E[X_{s+c} X_{t+c}]$.
\end{itemize}
\end{myDef}

It is easy to see that a strongly stationary process with finite mean and variation is a weakly stationary process.

\subsubsection{Logarithmic return is stationary\label{LR}}

Denote $P_t$ is the price of a financial underlying at time $t$, and $p_t$ is the logarithmic price, i.e. $p_t = \log{P_t}$. For any positive $a$, $\Delta a_t = p_t-p_{t-a}$ is the logarithmic return for the period $a$.

In Ref. \cite{12}, the author indicates that one can NOT REJECT the null hypotheses that logarithmic return time series are weakly stationary via statistical tests. Although the real hypothesis is that logarithmic return time series are strongly stationary, the gap between the theory and statistical tests is not very critical in application. So we will follow this conclusion. The test for stationarity will mentioned in section \ref{ADF}.

Under the premise that the logarithmic return $\{\Delta p_t\}$ is stationary, just follow the statements in Ref. \cite{Z.D.Wang}, when $N$ is large enough, one can assume that there is no more than one transaction in each closed time interval $\left[\frac{j}{N}, \frac{j+1}{N}\right]$.

$$\left(P_{\frac{j+1}{N}}/P_{\frac{j}{N}}\right)^{I_{\frac{j}{N}}} = \exp\left\{I_{\frac{j}{N}} \left(p_{\frac{j+1}{N}}-p_{\frac{j}{N}}\right)\right\}.$$

$I_n$ is the position on time $n$. $I = 1$ represents a long position, $I = -1$ represents a short position, and $I = 0$ represents a neutral position. Therefore the total realized logarithmic return (without counting transaction costs) by time $T$ is

$$\sum_{j<NT-1}{I_{\frac{j}{N}} \left(p_{\frac{j+1}{N}}-p_{\frac{j}{N}}\right)}.$$

In the China Financial Futures Exchange, the transaction costs of some products are at a small percentage of total value of transaction. Suppose the unilateral transaction cost percentage is $c$, the total realized logarithmic return by time $T$ is

$$\sum_{j<NT-1}{\left[I_{\frac{j}{N}} \left(p_{\frac{j+1}{N}}-p_{\frac{j}{N}}\right)+\log{\frac{1-c}{1+c}}\cdot\left|I_{\frac{j+1}{N}}- I_{\frac{j}{N}}\right|\right]}.$$

When $N\rightarrow\infty$, the above sum converges to

\begin{equation}
\int_0^T{I_s dp_s+ \log{\frac{1-c}{1+c}}K_T}= \sum_{i; T(i)<T}\left(p_{T_i}-p_{S_i}\right)\left(I_{T_i}-I_{S_i}\right)+\log{\frac{1-c}{1+c}}K_T.
\label{Eq1}
\end{equation}

$K_T$ is the number of transactions before $T$ and $S_i$ is the open time of the $i_{th}$ trade while $T_i$ is the close time of the $i_{th}$ trade.

When $\{(I_t, \Delta p_t)\}$ is stationary, so is each term in the sum of eq.\ref{Eq1}. Thus the mean logarithmic return as follows converges applying the strong ergodic theorem.

\begin{equation*}
\frac{1}{T} \left[\sum_{\{i; T(i)<T\}}\left(p_{T_i}-p_{S_i}\right)\left(I_{T_i}-I_{S_i}\right)+\log{\frac{1-c}{1+c}}K_T\right].
\end{equation*}

\subsubsection{Stationary moving average\label{SMA}}

According to Ref. \cite{Z.D.Wang}, neither $P_t$ nor $MA_t$ is stationary. Denote an n-period stationary moving average as follows,

\begin{equation}\nonumber
\begin{split}
SMA(n,P)_t = & MA(n,P)_t/P_t \\
=& \left(1+P_{t-1}/P_t+P_{t-2}/P_t+\cdots+P_{t-n+1}/P_t\right)/n\\
=& \left(1+e^{p_{t-1}-p_t}+e^{p_{t-2}-p_t}+\cdots +e^{p_{t-n+1}-p_t}\right)/n.
\end{split}
\end{equation}

An n-period stationary moving average is a function of the logarithmic return $\{\Delta p_t\}$. Moreover, the ratio of $SMA(n_s,P)_t$ and $SMA(n_l,P)_t$, denote $R(n_s,n_l,P)_t$, is also a function of $\{\Delta p_t\}$. When $\{\Delta p_t\}$ is stationary, so is $SMA(n,P)_t$ and the ratio.

$$R(n_s,n_l,P)_t = SMA(n_s,P)_t/SMA(n_l,P)_t = MA(n_s,P)_t/MA(n_l,P)_t.$$

Then $MA(n_s,P)_t$ up-crosses(down-crosses) $MA(n_l,P)_t$ is equivalent to $R(n_s,n_l,P)_t$ up-crosses(down-crosses) 1. As section \ref{MA} mentioned, we also set up a filter zone with an up-band and a low-band, in which no trading signals will be generated. So the trading rules with SMA and filter zone is defined as follows,

\begin{itemize}
  \item Open long: $R(n_s,n_l,P)_{t-1}\leq$1+up-band \& $R(n_s,n_l,P)_{t}>$1+up-band;
  \item Close long: $R(n_s,n_l,P)_{t-1}>$1+up-band \& $R(n_s,n_l,P)_{t}\leq$1+up-band;
  \item Open short: $R(n_s,n_l,P)_{t-1}\geq$1+low-band \& $R(n_s,n_l,P)_{t}<$1+low-band;
  \item Close short: $R(n_s,n_l,P)_{t-1}<$1+low-band \& $R(n_s,n_l,P)_{t}\geq$1+low-band.
\end{itemize}

Let low-band be the opposite number of up-band, then they can be described in one parameter $b$. Hence, we have the set of parameters $(n_s,n_l,b)$. The alternative values of $n_s$ are 1, 5, 10 and 15. The values of $n_l$ can be 20, 30, 60 and 120. b is chosen from $0.1\times 10^{-3}$, $0.5\times 10^{-3}$, $1\times 10^{-3}$ and $1.5\times 10^{-3}$. These parameters result in 64 stationary MA rules. Since we test 3 different frequency data, there 192 stationary MA strategies in total.

\subsubsection{Stationary $KDJ$\label{SKDJ}}

According to the definition of $KDJ$,

\begin{equation*}
\begin{split}
RSV_t = &100\times\frac{e^{p_t}-e^{\min\{p_t,p_{t-1},\cdots,p_{t-n+1}\}}}{e^{\max\{p_t,p_{t-1},\cdots,p_{t-n+1}\}}-e^{\min\{p_t,p_{t-1},\cdots,p_{t-n+1}\}}}\\
=&100\times\frac{e^{p_t-p_{t-n}}-e^{\min\{p_t-p_{t-n},p_{t-1}-p_{t-n},\cdots,p_{t-n+1}-p_{t-n}\}}}{e^{\max\{p_t-p_{t-n},p_{t-1}-p_{t-n},\cdots,p_{t-n+1}-p{t-n}\}}-e^{\min\{p_t-p_{t-n},p_{t-1}-p_{t-n},\cdots,p_{t-n+1}-p_{t-n}\}}}.
\end{split}
\end{equation*}

When $\{\Delta p_t\}$ is stationary, so is $\{RSV_t\}$. $\%K$ is the function of $RSV$ while $\%D$ is the function of $\%K$. Therefore $\{\%K_t\}$ and $\{\%D_t\}$ are stationary if $\{\Delta p_t\}$ is stationary.

As for the trading strategy, we choose a strategy as follows, which is very popular among traders.

\begin{itemize}
  \item Open long: $\%K_{t-1}<\%D_{t-1}$ \& $\%K_t\geq \%D_t$ \& $20\leq \%K_t\leq 80$;
  \item Close long: $\%K_{t-1}>\%D_{t-1}$ \& $\%K_t\leq \%D_t$;
  \item Open short: $\%K_{t-1}>\%D_{t-1}$ \& $\%K_t\leq \%D_t$ \& $20\leq \%K_t\leq 80$;
  \item Close short: $\%K_{t-1}<\%D_{t-1}$ \& $\%K_t\geq \%D_t$.
\end{itemize}

There are 3 parameters for $KDJ(m,n,k)$. For each high-frequency data, we pick parameters from (5,1,3), (5,3,3), (9,3,3), (14,3,3) and (19,3,3), resulting in 15 $KDJ$ strategies.

\subsubsection{Stationary Bollinger {red}band\label{SBoll}}

Obviously, the original Bollinger bands is not stationary. We can transform Bollinger bands into the following form.

\begin{equation}
\begin{split}
SBoll(n,P)_t = & \frac{P_t-EMA(n,P)_t}{\sigma_t}\\
=&\frac{P_t-\frac{nP_t+(n-1)P_{t-1}+\cdots+P_{t-n+1}}{n(n+1)/2}}{\sqrt{\frac{1}{n-1}\sum_{i=t-n+1}^{t}(P_i-EMA(n,P)_t)^2}}\\
=&\frac{e^{p_t}-\frac{2}{n(n+1)}(ne^{p_t}+(n-1)e^{p_{t-1}}+\cdots+e^{p_{t-n+1}})}
{\sqrt{\frac{1}{n-1}[e^{p_t}-\frac{2}{n(n+1)}(ne^{p_t}+\cdots+e^{p_{t-n+1}})]^2+\cdots+\frac{1}{n-1}[e^{p_{t-n+1}}-\frac{2}{n(n+1)}(ne^{p_t}+\cdots+e^{p_{t-n+1}})]^2}}\\
=&\frac{e^{p_t-p_{t-n}}-\frac{2}{n(n+1)}(ne^{p_t-p_{t-n}}+(n-1)e^{p_{t-1}-p_{t-n}}+\cdots+e^{p_{t-n+1}-p_{t-n}})}
{\sqrt{\frac{1}{n-1}\sum_{i=t-n+1}^{t}[e^{p_i-p_{t-n}}-\frac{2}{n(n+1)}(ne^{p_t-p_{t-n}}+\cdots+e^{p_{t-n+1}-p_{t-n}})]^2}}.
\end{split}
\end{equation}

$SBoll(n,P)_t$ is a function of $\{e^{p_{t-i}-p_{t-n}}; i=1,2,\cdots,n-1\}$. When $\{p_t\}$ is of strongly stationary increment, $\{SBoll(n,P)_t\}$ is a stationary process. The price between the Bollinger bands is equivalent to the inequality $-K<SBoll(n,P)_t<K$.

So the trading signals are generated by the following rules:
\begin{itemize}
  \item Open long: $SBoll(n,P)_{t-1}\leq K$ \& $SBoll(n,P)_t > K$;
  \item Close long: $SBoll(n,P)_{t-1}\geq K$ \& $SBoll(n,P)_t < K$;
  \item Open short: $SBoll(n,P)_{t-1}\geq -K$ \& $SBoll(n,P)_t < -K$;
  \item Close short: $SBoll(n,P)_{t-1}\leq -K$ \& $SBoll(n,P)_t > -K$.
\end{itemize}

Trading signals generated by stationary Bollinger bands strategy are associated with 2 parameters. One for $EMA(n,P)_t$, the other is the coefficient of $\sigma$. The value of n can be 20, 30, 60 and 120, while K is chosen from 0.1, 0.5, 1, 1.5, 2, and 2.5. Therefore, there are 72 stationary Bollinger bands strategies with 3 high-frequency data series.

All high-frequency stationary strategies above just trade no more than one share in a unit of time and will be forced to close all positions at the end of every trading day.

\section{Data description\label{Data}}

We apply the trading rules in above strategies on CSI300(China Security Index 300) Stock Index Futures main contract to investigate which technical strategy has the best performance. The reasons why we choose this underlying asset are as follows.

CSI300 Stock Index Futures is the first stock index future in Chinese financial market introduced by the China Financial Futures Exchange(CFFEX) on April 16th 2010. CSI300 consists of 300 A shares in Shanghai and Shenzhen securities markets. The sample covers about sixty percent of the market value of Shanghai and Shenzhen, exhibiting good representativeness. Hence, CSI300 Stock Index Futures has great market liquidity and reflects the overall trends of the market. The other advantage of choosing CSI300 Stock Index Futures is that although short selling is forbidden in Chinese A share security market, it is allowed in futures market. This will bring convenience and completeness to our research.

\begin{figure}[H]
\centering
\includegraphics[width=12cm,height=9cm]{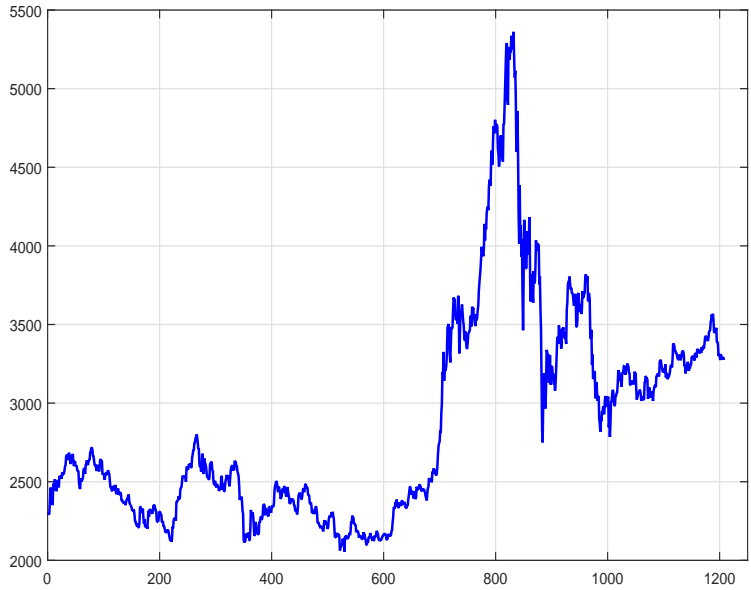}
\caption{Daily data of IF main contract from 20120104 to 20161230}
\label{FIG1}
\end{figure}

To the comprehensiveness of our research, we choose 3 different high-frequency data(15s,30s and 60s) to test all the trading rules. All three high-frequency data of CSI300 Stock Index Futures cover a period from January 4th 2012 to December 30th 2016, which leads to 1276799(15s), 638399(30s) and 319199(60s) data points respectively. Due to the large data volume, one can hardly distinguish the difference of high-frequency data and daily data in a figure. Thus, Fig. \ref{FIG1} just shows the trend of daily data. One thing should be mentioned is that before January 1st 2016 CSI300 Stock Index Futures could be traded from 9:15 am to 11:30 am and from 13:00 pm to 15:15 pm each trading day. However, the open time changed to 9:30 and the close time changed to 15:00 from January 1st 2016. The data we used follows this rule.

\begin{table}[H]
\caption{Statistical description of CSI300 Stock Index Future}
\begin{tabular}{llllllll}
\hline
Frequency & Observation & Max & Min & Mean & Std. & Kurtosis& Skewness \\
\hline
15s & 1276799 & 8.0172\% & -6.3485\% & 2.5267E-07 & 0.0006 & 986.0586 & 0.7322 \\
30s & 638399 & 7.8433\% & -5.7044\% & 5.0467E-07 & 0.0008 & 512.5485 & 1.3490 \\
60s & 319199 & 7.1733\% & -5.8385\% & 1.0122E-06 & 0.0012 & 208.6592 & 0.9192 \\
\hline
\end{tabular}
\label{TAB1}
\end{table}

We estimate the log-return, i.e. logarithmic difference return, of three high-frequency data separately, and find that the largest log-return of 15s, 30s and 60s data are 8.02\%, 7.84\% and 7.71\% respectively while the smallest log-return of 15s, 30s and 60s data are -6.35\%, -5.70\% and -5.84\%. Table \ref{TAB1} lists the basic statistics of log-returns for all three high-frequency data. When compared to a normal distribution, all three return distributions have excess kurtosis and right skewness.

\section{Methodology\label{Section3}}

\subsection{Performance measure}

\subsubsection{Performance and annual performance\label{AR}}

The performance of a strategy is defined by how much the net profit or loss the strategy generated based on the initial equity at the end of given period of time or in other word, the back test period. Different strategies may have different back test periods, for the convenience to compare the performance of different strategies, annual performance is a better choice. Annual performance calculates a performance over a year, which equals to a performance multiplied by 250(number of trading days in a year) and divide the number of days in the back test period.

\subsubsection{PnL (Profit and Loss)\label{PnL}}

Profit and loss index, in short, $PnL$ is the ratio of ¡°net profit¡± over the amount of winning or losing trades. It is defined by the following formula.

$$PnL = \frac{Trade Profit-Trade Loss}{Max(Trade Profit, Trade Loss)}$$

The value of $PnL$ ranges from -100 to +100. A positive number reveals that overall a strategy generate a net profit, otherwise a net loss.

\subsubsection{Average profit/Average loss\label{AP/AL}}

This index is a ratio of average profit from profitable trades over average loss from unprofitable ones. If a strategy is ¡°good¡±, it will let profitable trade run and stop loss as quickly as possible. That is to say, the ratio is high if one strategy is ¡°good¡± in above meaning.

In consideration of the convenience of horizontal comparison, we will calculate the average profit and loss via daily return in this paper.

\subsubsection{Percentage of profitable trade\label{WR}}

This index gives the win rate of a strategy, which is the ratio of the number of profitable trades over whole trade number. A high ratio indicates the strategy has a high probability to predict the change direction of price correctly.

A profitable strategy, however, may not have a high win rate. For example, if there are 4 trades. A profitable one with 5 points, and three loss ones with 1 point each. The win rate is only 25\%, but the total net is 2 points.

\subsection{Risk measure}

\subsubsection{Sharp ratio\label{SR}}

Sharp ratio is an index estimating how much excess return a strategy can capture via a unit of risk. The excess return is the difference between return of the strategy and the risk-free interest rate, and the standard deviation of return generated by the strategy is considered to be the risk. Specific formula is as follows,

$$Sharp \ ratio =\frac{E(r)-r_f}{\sqrt{Var(r)}}.$$

$r_f$ is the risk-free interest and $r$ is the return of measured strategy.

Deserve to be mentioned, we calculate daily sharp ratio then convert it to annual sharp ratio to measure our strategies, since high-frequency sharp ratio is not representative enough for comparison while calculate annual sharp ratio directly need more than 5 years data to be meaningful. $r_f$ will be set to be zero when calculate daily sharp ratio since the forced liquidation at the end of every trading day. The converting formula is as follows,

$$Annual \ Sharp \ ratio = Daily \ Sharp \ ratio\times\sqrt{250}$$

The number 250 is the average number of trading days in one year.

\subsubsection{Maximum drawback percentage\label{MDP}}

Maximum drawback percentage is another popular index to measure risk. It presents the worst situation of a strategy in a given period of time. Suppose there are n trading days. Denote $V_i$ is the asset value on the $i_{th}$ day and $W_i = \max_{j = 1,2,\ldots,i}\{V_j\}$

$$Maximum\ drawback\ percentage = \max_{i = 1,2,\ldots,n}\left\{1-\frac{V_i}{W_i}\right\}$$

The smaller this index is, the robuster the strategy is. We also concentrate on maximum drawback percentage based on daily return.

\subsubsection{Annual return / Maximum drawback percentage\label{AR/MDP}}

Just as the name implies, this index is the ratio of annual return over maximum drawback percentage. This index is popular and important comprehensive standard to measure the stability of one strategy.

\subsection{Testing statistics}

\subsubsection{ADF-test\label{ADF}}

ADF-test is the abbreviation of augmented Dickey-Fuller test. Although this test can only test a unit root process, not a weakly stationary process, the similarity between them gives the ADF-test the ability to check the weakly stationary process. Many papers have been published on the ADF-test. The Dickey-Fuller test \cite{Dickey1,Dickey2} is a popular unit root test used to assess the time-series property of economic and financial data. MacKinnon \cite{MacKinnon}, Harris \cite{Harris} and Cheung \& Lai \cite{Cheung} focused on the use of test parameters, therein demonstrating that both the lag order and the sample size can affect the finite-sample behavior of the test.

The ADF-test for a unit root assesses the null hypothesis of a unit root using the model $y_{t}=c+\delta t+\phi y_{t-1}+\beta_{1}\Delta y_{T-1}+\beta_{2}\Delta y_{T-2}+\cdot\cdot\cdot+\beta_{p}\Delta y_{t-p}+\epsilon_{t}$, where $\Delta$ is the differencing operator, $p$ is the number of lagged difference terms, and $\epsilon_{t}$ is a mean zero innovation process. The null hypothesis of a unit is $H_{0} : \phi=1$; under the alternative hypothesis, $\phi<1$. Variants of the model allow for different growth characteristics. The model with $\delta=0$ has no trend component, and the model with $c=0$ and $\delta=0$ has no drift or trend. The logic applied by statisticians is as follows: if one cannot show that the current sample path is unlikely from a weakly stationary process, then we simply keep the weakly stationarity hypothesis.

We will conduct ADF-test to verify the stationarity of logarithmic return of high-frequency data.

\subsubsection{Superior predictive ability\label{SPA}}

As White \cite{White} mentions, when a given data set is reused more than once, some strategies may be significant just by chance rather than any inherent merit. This phenomenon is called data snooping bias. White \cite{White} creates the reality check(WRC) test to correct the data snooping effect. Hansen \cite{Hansen} proposes superior predictive ability(SPA) test which is modified from WRC and more powerful under most circumstances. Thus, we choose SPA test to check data snooping bias.

Assume there are $K$ different strategies and $T$ trading days, denote $\overline{d_{k,t}} = \sum_{t = 1}^T d_{k,t}$ as the performance of strategy $k$, where $d_{k,t}(k \in {1, 2, \ldots, K},t \in {1, 2, \ldots, T} )$ represents the performance measure of methods $k$ on day $t$. Suppose there are $N$ time points on day $t$, and the unilateral transaction cost is $c$, the performance measure $d_{k,t}$ is written as,

$$d_{k,t} = \sum_{n = 1}^{N-1}\left(\log{\frac{P_{n+1}}{P_n}}\cdot I_n+\log{\frac{1-c}{1+c}}\cdot|I_{n+1}-I_n|\right)=\sum_{n = 1}^{N-1}\left(\Delta{p_{n+1}}\cdot I_n+\log{\frac{1-c}{1+c}}\cdot|I_{n+1}-I_n|\right)$$

$I_n$ is the position on time $n$. $I = 1$ represents a long position, $I = -1$ represents a short position, and $I = 0$ represents a neutral position.

Denote $\mu_k = E(d_{k,t})$. The null hypothesis of SPA test is as followed:
$$H_0 : \mu_k \leq 0; k \in {1, 2, \ldots, K}$$

Politis and Romano \cite{Politis} proposes a stationary bootstrap, we follow this method to get SPA test result. First, generate a random matrix $\{R_{b,t}\}$ of size $B\times T$. For each row, for example the $b_{th}$ row, $P^*(R_{b,1}=t,t = 1,2,\ldots,T) = 1/T$. When $t>1$, $P^*(R_{b,t} = R_{b,t-1}+1) = Q$ and $P^*(R_{b,t} = s,s = 1,2,\ldots,T)=(1-Q)/T$.

Second, denote $\{d^*_{k,t}(b)\triangleq d^*_{k,R_{b,t}}, t = 1,2,\ldots,T\}$ as the $b_{th}$ re-sample. The performance of the $b_{th}$ re-sampled is $\overline{d^*_k}(b) = \sum^T_{t = 1}d^*_{k,t}(b) /T$. Denote $P^*$ as the bootstrap probability measure, the critical value for SPA test under significance level of $\alpha$ can be described as

$$\hat{q^*_\alpha} = \max(0,\hat{q_\alpha}), \  \  \hat{q_\alpha} = \inf{\{q|P^*[\sqrt{n}\max_{k=1,2,\ldots,K}(\overline{d^*_k}-\overline{d_k}+\hat{\mu_k})/\hat{\omega_k}\leq q]\geq 1-\alpha\}}$$

$\hat{\mu_k} = \overline{d_k} \chi\{ \sqrt{n}\overline{d_k} \leq -\hat{\omega_k}\sqrt{2\log\log n}\}$, and $\hat{\omega_k}$ represents the consistent estimator of $\omega_k \triangleq var( \sqrt{n}\overline{d_k})$. Follow Hansen's \cite{Hansen} recommendation, the kernel estimator of $\hat{\omega_k}$ is used in this paper. The null hypothesis is rejected if $\max_{k = 1,2,\ldots,K}\sqrt{n}\overline{d_k}/\hat{\omega_k}>\hat{q^*_a}$.

With further development, P.H.Hsu, et al. \cite{P.H.Hsu} extend SPA test to Step-SPA test which enables researchers to identify significant models. Step-SPA test can be conducted through the following steps:

\begin{itemize}
  \item Step 1: Re-arrange ${\overline{d_k}/\hat{\omega_k},k = 1,2,\ldots,K}$ in a descending order.
  \item Step 2: Assume the $k_{th}$ model has the max $\overline{d_k}/\hat{\omega_k}$. Procedure stops if $\sqrt{n}\overline{d_k}/\hat{\omega_k}>\hat{q^*_a}$, otherwise move to step 3.
  \item Step 3: Remove the $k_{th}$ model in step 2 from the model universe, then conduct step 1 and step 2 for the remaining models.
  \item Step 4: Repeat the 3rd step till no model can be rejected. All the removed models are identified as significant ones.
\end{itemize}

Following P.H.Hsu, et al. \cite{P.H.Hsu}, we let $Q = 0.9$ and $B = 500$, and set the significance level to 5\% and 10\%.

\section{Empirical results\label{Section4}}

\subsection{Performance\label{Performance}}

We set initial capital to be one million and trade no more than one share once.
We test 64 groups parameters of stationary MA on 3 high-frequency data respectively, resulting in 192 different strategies. If parameters for short MA and long MA are fixed, number of trades has a decreasing trend while width of filter band rises. For each frequency data, more than four-fifth strategies have a positive annual return, the specific numbers are 48 of 64, 52 of 64 and 56 of 64. However, strategies have sharp ratio larger than 1.5 are not that many, only 30 in total. Another thing should be mentioned is that only 16 strategies win more than 50\%. Stationary MA strategy is a trend-following strategy which leads to a low win rate. Wrong judgement of the direction of trend or repeated shocks may result in several losses, however, capture megatrends once will eave the situation even bring some profit.

Popular parameters for KDJ is not such many as for MA. Here are 15 KDJ strategies together. 15s KDJ strategies do not perform well, none make profit, while three of five 30s strategies and all five 60s strategies have positive annual return.

As for stationary Bollinger bands strategies, 24 of 72 make profit, and the best one earns 59.31\% one year. However, only 3 strategies have sharp ratio larger than 1.5 and only 1 strategy win more than 50\%. Not high enough sharp ratio indicates that the stability should be strengthened. Trading rules generated by Bollinger bands belong to channel breakouts strategy. Similar to trend-following strategy, it makes money depending on high ratio of profit over loss instead of high win rate. To be mentioned, ratios of average profit over average loss of all 279 strategies are all larger than 1.

\begin{table}[htbp]\scalebox{0.85}{
  \centering
  \begin{threeparttable}
  \caption{Performance of 3 strategies}\label{TAB3}
    \begin{tabular}{lrrrrrrrrrrr}
    \toprule
    Startegy & \multicolumn{1}{l}{LDT\tnote{1}}& \multicolumn{1}{l}{SDT\tnote{1}} & \multicolumn{1}{l}{ASP\tnote{2}} & \multicolumn{1}{l}{ADP\tnote{2}} & \multicolumn{1}{l}{AR\tnote{3}} & \multicolumn{1}{l}{MDP\tnote{3}} & \multicolumn{1}{l}{AR/MDP\tnote{3}} & \multicolumn{1}{l}{SR\tnote{4}} & \multicolumn{1}{l}{PnL\tnote{4}} & \multicolumn{1}{l}{WR\tnote{5}} & \multicolumn{1}{l}{AP/AL\tnote{5}} \\
    \midrule
    MA\_15(10,120,0.0001) & 10353 & 10384 & 113.90 & 1953.70 & 48.84\% & 6.33\% & 7.71  & 2.17  & 0.36  & 51.94\%  & 1.45 \\
    KDJ\_60(5,1,3) & 20035 & 19648 & 41.31 & 1355.88 & 33.90\% & 11.68\% & 2.90  & 2.24  & 0.38  & 51.94\%& 1.49 \\
    Boll\_30(120,0.1) & 22234 & 22325 & 64.37 & 2372.41 & 59.31\% & 6.64\% & 8.93  & 2.33  & 0.40  & 50.79\%  & 1.62 \\
    \bottomrule
    \end{tabular}%
    {\tiny
    \begin{tablenotes}
    \footnotesize
    \item[1] LDT is the abbreviation of long deal times and SDT is the abbreviation of short deal times
    \item[2] ASP means average single profit and ADP means average daily profit
    \item[3] AR stands for annual return, MDP means maximum drawback percentage and AR/MDP is the ratio of them
    \item[4] SR is sharp ratio index and PnL is Profit and Loss index
    \item[5] WR means win rate and AP/AL is the ratio of average profit over average loss
    \end{tablenotes}}
  \end{threeparttable}}
\end{table}%

\begin{figure}[H]
\centering
\includegraphics[width=12cm,height=9cm]{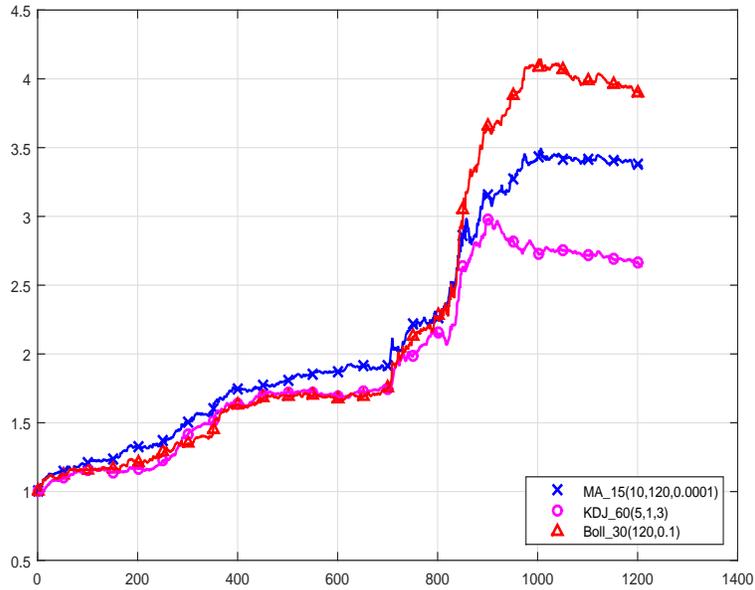}
\caption{Accumulated return of 3 strategies from 20120104 to 20161230}
\label{FIG2}
\end{figure}

The most profitable strategy of each technical indicator strategy is listed in Table \ref{TAB3} separately, and Fig. \ref{FIG2} shows the net value of these 3 strategies. For the overall simplicity, detail results of all 279 strategies are presented in Table \ref{TAB:MA1}-\ref{TAB:Boll3} in appendix.

Above shows the circumstance when transaction cost is not taken into account. Once the cost is included, no strategies can bring profit. There are two possible reasons. As Table \ref{TAB2} shows, the transaction cost has been adjusted higher since the next half year of 2015, especially the adjustment in September 7th 2015, which leaded the new cost to be hundredfold of the cost from the next half year of 2012 to the first half year of 2015. This change does not only increase the cost but also fluctuates the market pattern. The good news is that the transaction cost has been adjusted to 9\%\% which gives investors the hope that one day it will be lower in the future. The other reason to explain this result is the market efficiency. There's a bold assumption that the transaction cost may has the function to insure the efficiency of market. It worth noting that in this paper we just discuss the simple stationary technical trading rules. They are basic rules and has the potential of further development. Just as strategies given by Ref. \cite{Liu,X.Wang,Bao} show, there are methods to cover the cost and make a profit. However, this is not the area of this paper.

\begin{table}[H]
\caption{Transaction costs from 20120104 to 20161230}
\begin{tabular}{lllllll}
\hline
Start Date & 20120104 & 20120601 & 20120901 & 20150803 & 20150826 & 20150907 \\
\hline
Cost & 0.5\%\% & 0.35\%\% & 0.25\%\% & 0.23\%\% & 1.15\%\% & 23\%\% \\
\hline
\end{tabular}
\label{TAB2}
\end{table}

\subsection{ADF-test}

\begin{figure}[H]
\centering
\subfigure[]{\includegraphics[width=5cm,height=3.75cm]{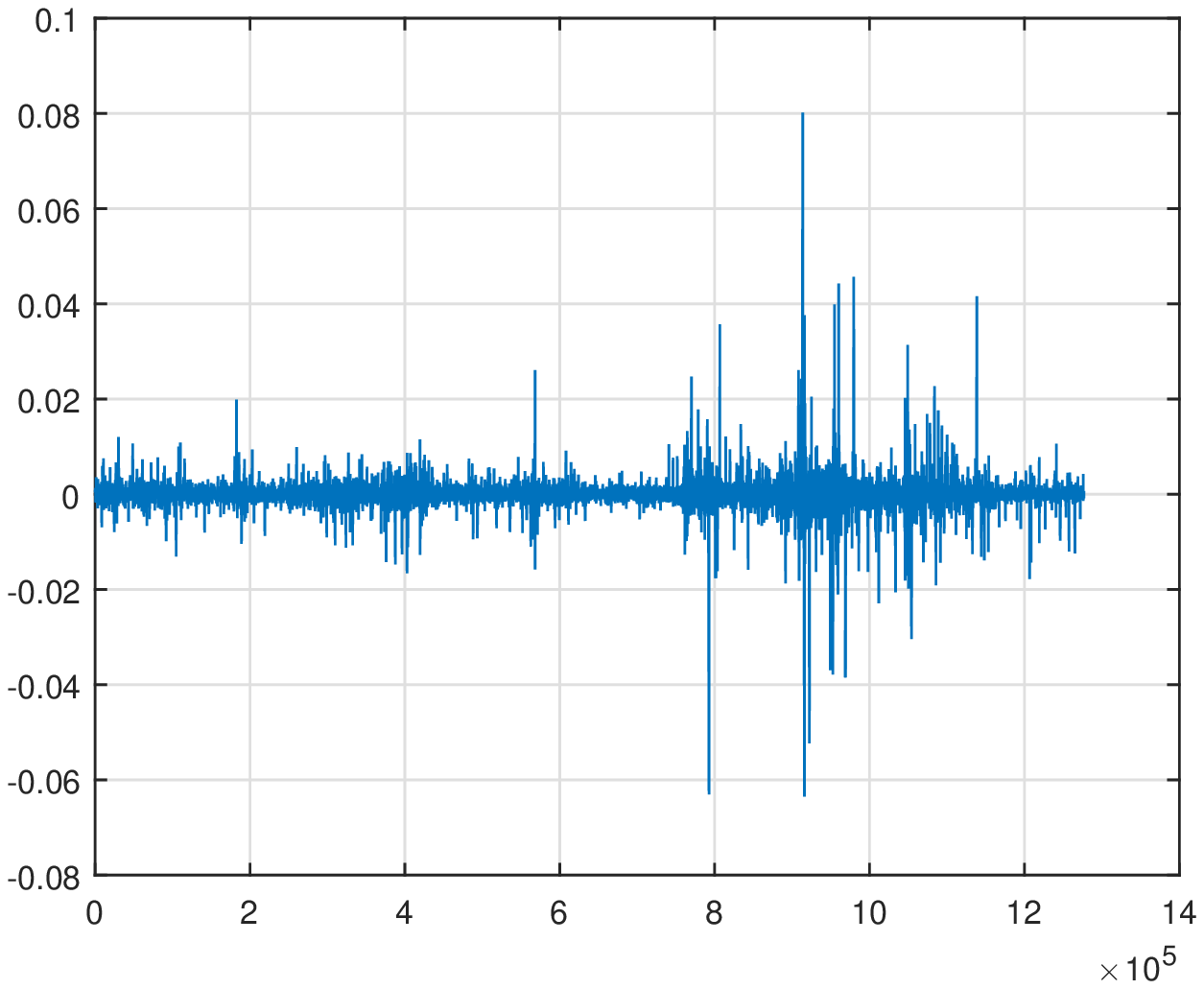}}
\subfigure[]{\includegraphics[width=5cm,height=3.75cm]{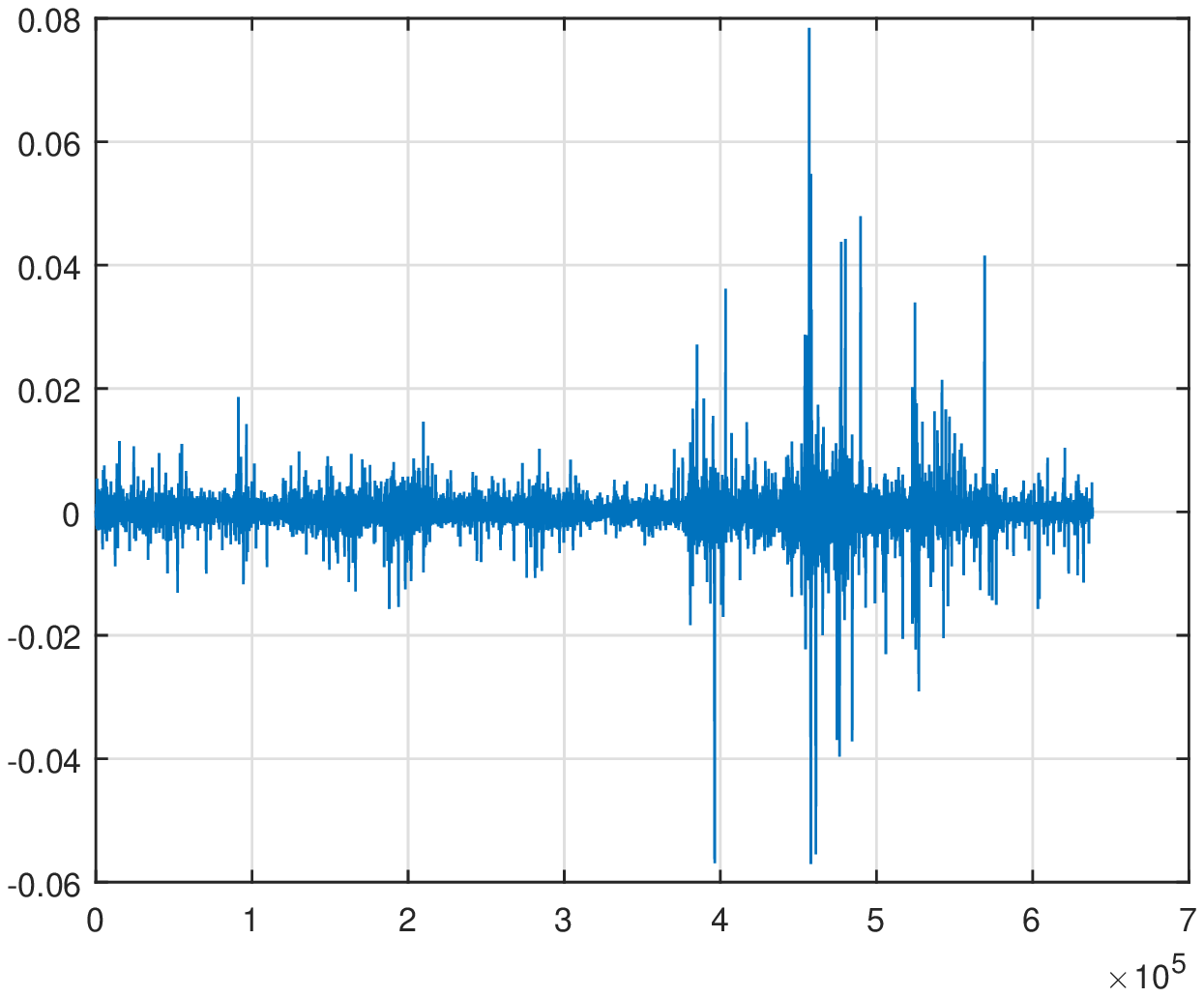}}
\subfigure[]{\includegraphics[width=5cm,height=3.75cm]{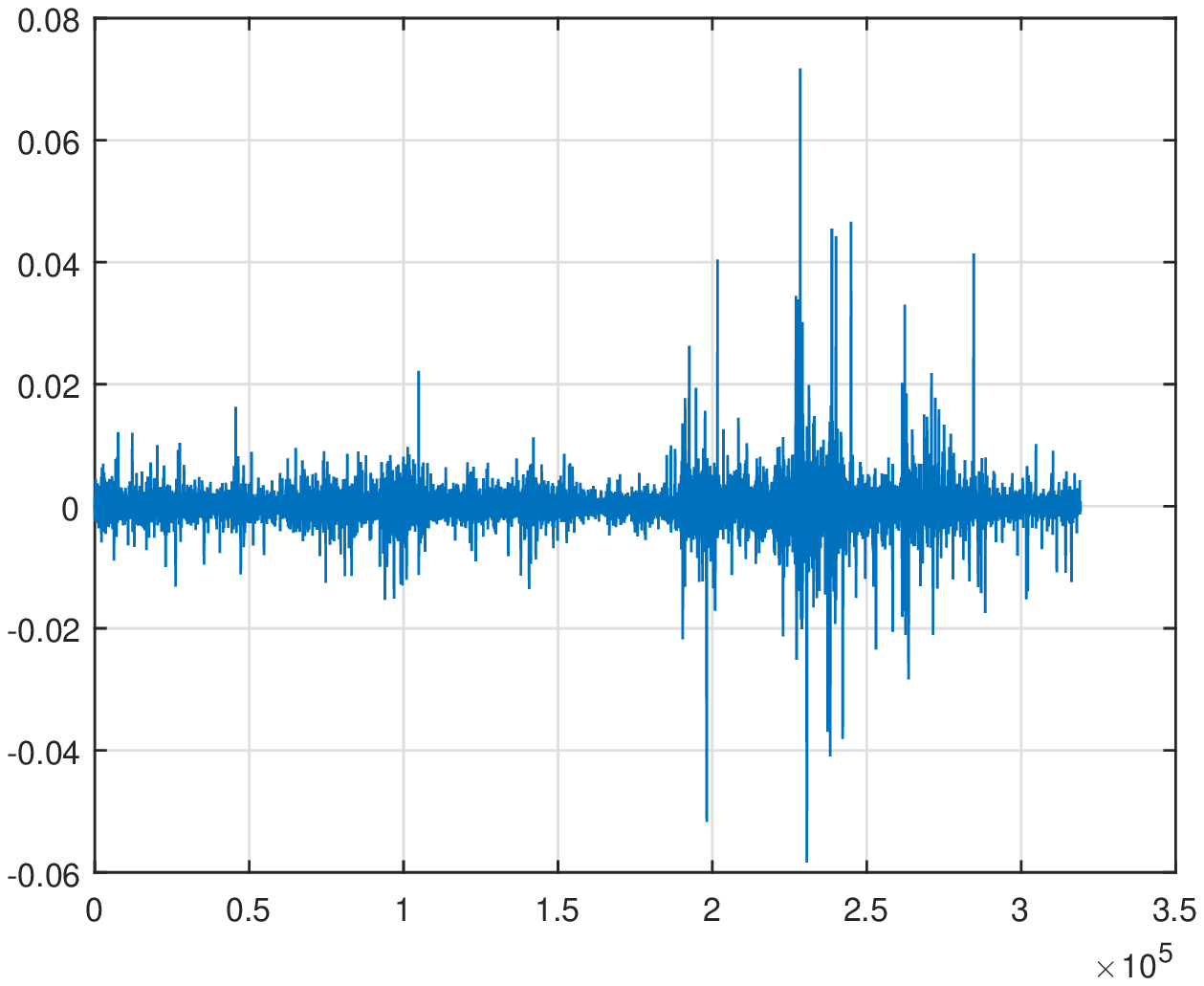}}
\caption{Log-return of IF main contract data from 20120104 to 20161230}
\label{FIG3}
\end{figure}

First, we test the stationarity of log-return of raw time series, i.e. $\{\Delta p_t\}$. Tables \ref{TAB41} \ref{TAB42} \ref{TAB43} are simplified 15s,30s and 60s data test results. Regardless of the number of lagged difference terms ($lags$) that are taken, the final test result remains ``1'', which means that the logarithmic return is a weakly stationary process. On the other hand, in the Black-Scholes framework \cite{Z.D.Wang}, the stock price can be written as $P_{t}=P_{0}\exp\{\sigma B_{t} + (r-\sigma^{2}/2)t\}$; thus, $\Delta p_{t}=\sigma (B_{t}-B_{t-1})+(r-\sigma^{2}/2)$, where $B_{t}-B_{t-1}$ is the difference between two independent Brownian motions, and therefore, it is stationary. In summary, according to the classic financial models and sample data verification (ADF test), $\{\Delta p_{t}\}$ is a stationary process. Fig.\ref{FIG3}(a),(b)and(c)show the logarithmic return of 15s, 30s and 60s data respectively.


\begin{table}[H]
\centering
 \begin{threeparttable}
 \begin{tabular}{llll}
 \toprule
   & $ test-t1,\ lags-0$ & $test-t1,\ lags-1$ & $test-t1,\ lags-2$ \\
 \midrule
coeff  &     $-0.0115$ & $[-0.0286;0.0170]$     & $[-0.04;0.03;0.01]$\\
tStats &    $-12.9587$ & $[-22.7347;19.1567]$   & $[-0.0414;0.0296;0.0125]$\\
FStat  &         $Inf$ & $535.2234$             & $366.8779$      \\
AIC    & $-1.5532e+07$ & $-1.5532e+07$          & $-1.5532e+07$   \\
BIC    & $-1.5531e+07$ & $-1.5532e+07$          & $-1.5532e+07$   \\
p-value&    $1.00e-03$ & $1.00e-03$             & $1.00e-03$      \\
H      &           $1$ & $1$                    & $1$             \\
\bottomrule
\end{tabular}
\caption{ 15s ADF-test Result }
\label{TAB41}
\end{threeparttable}
\end{table}

\begin{table}[H]
\centering
\begin{threeparttable}
 \begin{tabular}{llll}
 \toprule
   & $ test-t1,\ lags-0$ & $test-t1,\ lags-1$ & $test-t1,\ lags-2$          \\
 \midrule
coeff  &     $-0.0310$  & $[-0.0134;-0.0171]$  & $[-0.0198;-0.0106;0.0063]$ \\
tStats &     $-24.8163$ & $[-7.4790;-13.6445]$ & $[-8.9914;-5.9158;4.9963]$ \\
FStat  &         $Inf$  & $802.4794$           & $413.7396$                 \\
AIC    & $-7.3326e+06$  & $-7.3327e+06$        & $-7.3328e+06$              \\
BIC    & $-7.3326e+06$  & $-7.3327e+06$        & $-7.3327e+06$              \\
p-value&    $1.00e-03$  & $1.00e-03$           & $1.00e-03$                 \\
H      &           $1$  & $1$                  & $1$                        \\
\bottomrule
\end{tabular}
\caption{ 30s ADF-test Result }
\label{TAB42}
\end{threeparttable}
\end{table}

\begin{table}[H]
\centering
\begin{threeparttable}
 \begin{tabular}{llll}
 \toprule
   & $ test-t1,\ lags-0$ & $test-t1,\ lags-1$ & $test-t1,\ lags-2$       \\
   \midrule
coeff  &     $0.0047$  & $[0.0134;-0.0088]$   & $[0.0090;-0.0044;0.0044]$\\
tStats &     $2.6326$  & $[5.3600;-4.9529]$   & $[2.9693;-1.7582;2.4835]$\\
FStat  &         $Inf$ & $31.7311$            & $18.9508$                \\
AIC    & $-3.4501e+06$ & $-3.4501e+06$        & $-3.4501e+06$            \\
BIC    & $-3.4501e+06$ & $-3.4501e+06$        & $-3.4501e+06$            \\
p-value&    $1.00e-03$ & $1.00e-03$           & $1.00e-03$               \\
H      &           $1$ & $1$                  & $1$                      \\
\bottomrule
\end{tabular}
\caption{ 60s ADF-test Result }
\label{TAB43}
\end{threeparttable}
\end{table}

\begin{figure}[H]
\centering
\subfigure[]{\includegraphics[width=8cm,height=6cm]{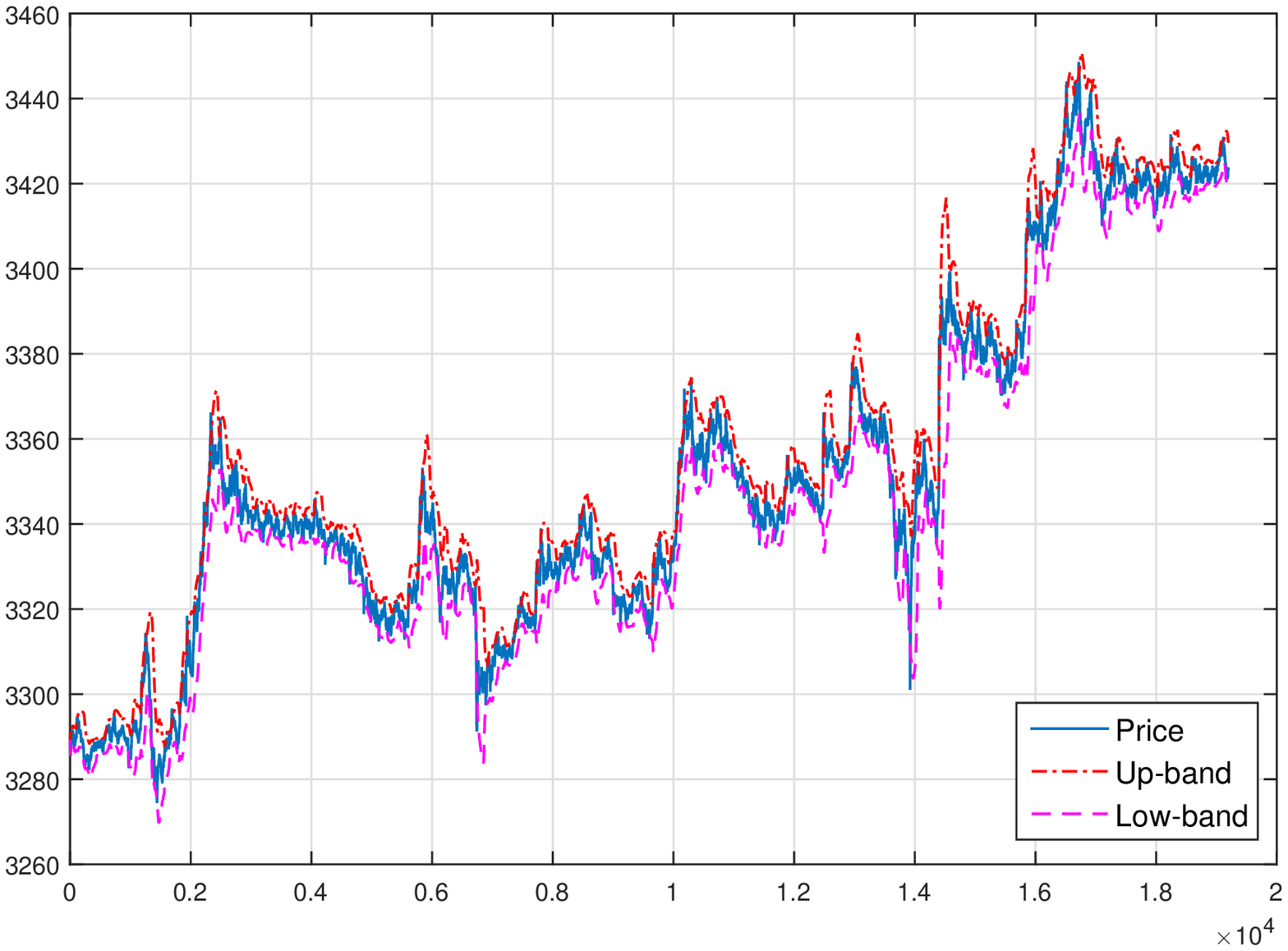}}
\subfigure[]{\includegraphics[width=8cm,height=6cm]{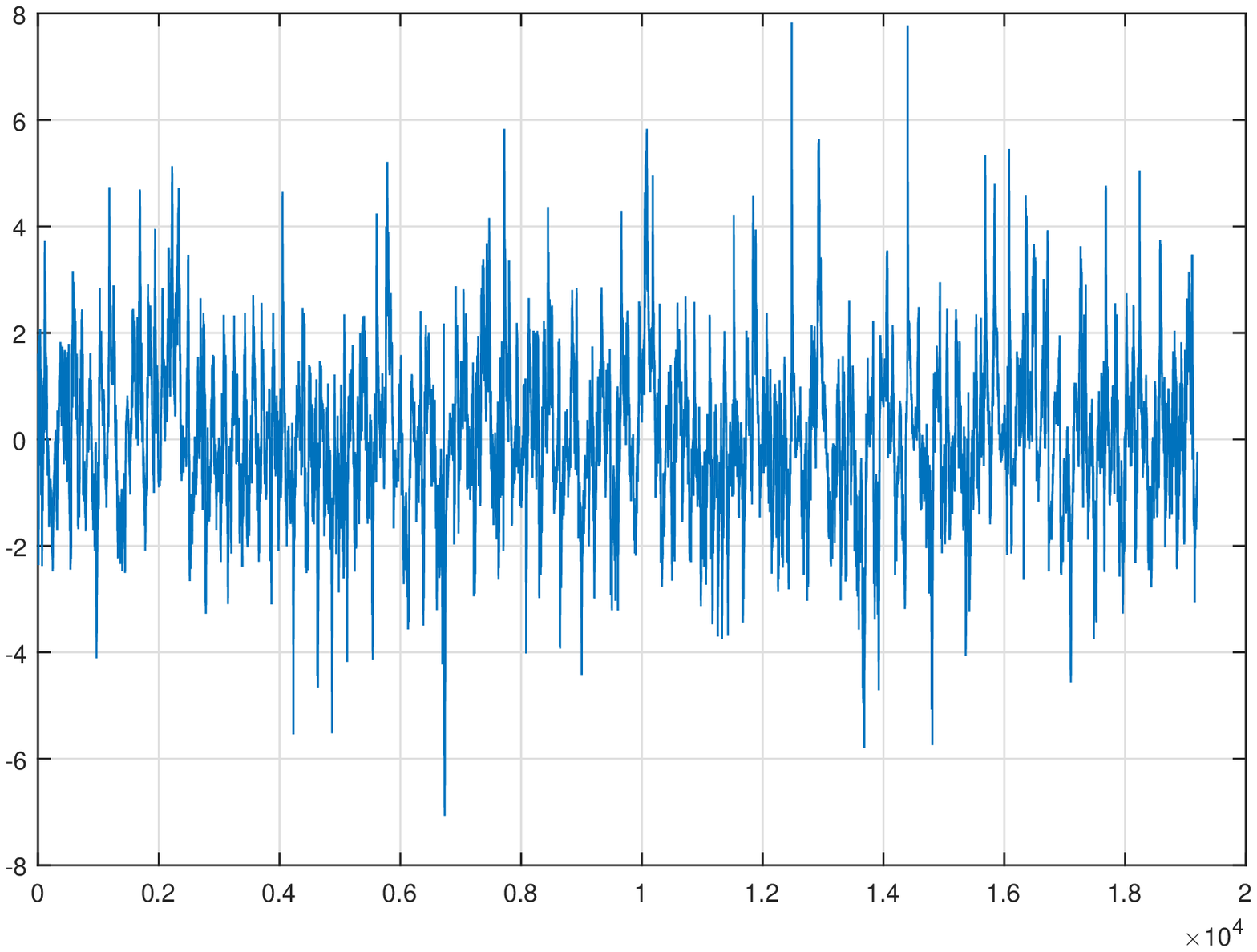}}
\caption{Comparison of original Bollinger bands and stationary Bollinger band with 15s data of IF1611}
\label{FIG:IF1611}
\end{figure}

Second, we verify the stationarity of three stationary indicators and find that all reject the null hypothesis. In other words, we can not reject that these indicators are stationary. For simplicity, here just show the result of $SBoll(120,2.5)$ with 15s data in Table \ref{TAB44}. Fig. \ref{FIG:IF1611} indicates the shape of $Boll(120,2.5)$ and $SBoll(120,2.5)$ with 15s data of IF1611 from 20th October 2016 to 16th November 2016. From Fig.\ref{FIG:IF1611}(a), original Bollinger's bands have clear tendency. However, stationary Bollinger band in Fig.\ref{FIG:IF1611}(b) fluctuates around 0.

\begin{table}[H]
\centering
\begin{threeparttable}
 \begin{tabular}{llll}
 \toprule
   & $ test-t1,\ lags-0$ & $test-t1,\ lags-1$ & $test-t1,\ lags-2$           \\
   \midrule
coeff  &     $0.9582$     & $[0.9607;-0.0607]$        & $[0.9621;-0.0643;-0.0356]$      \\
tStats &     $3.7799e+03$ & $[3.7571e+03;-68.6809]$   & $[3.7305e+03;-72.4007;-40.2039]$\\
FStat  &         $Inf$    & $1.4345e+07$              & $7.1826e+06$                    \\
AIC    & $1.3410e+06$     & $1.3363e+06$              & $1.3346e+06$                    \\
BIC    & $1.3410e+06$     & $1.3363e+06$              & $1.3347e+06$                    \\
p-value&    $1.00e-03$    & $1.00e-03$                & $1.00e-03$                      \\
H      &           $1$    & $1$                       & $1$                             \\
\bottomrule
\end{tabular}
\caption{ ADF-test result of 15s stationary Bollinger bands }
\label{TAB44}
\end{threeparttable}
\end{table}

\begin{figure}[H]
\centering
\subfigure[]{\includegraphics[width=8cm,height=6cm]{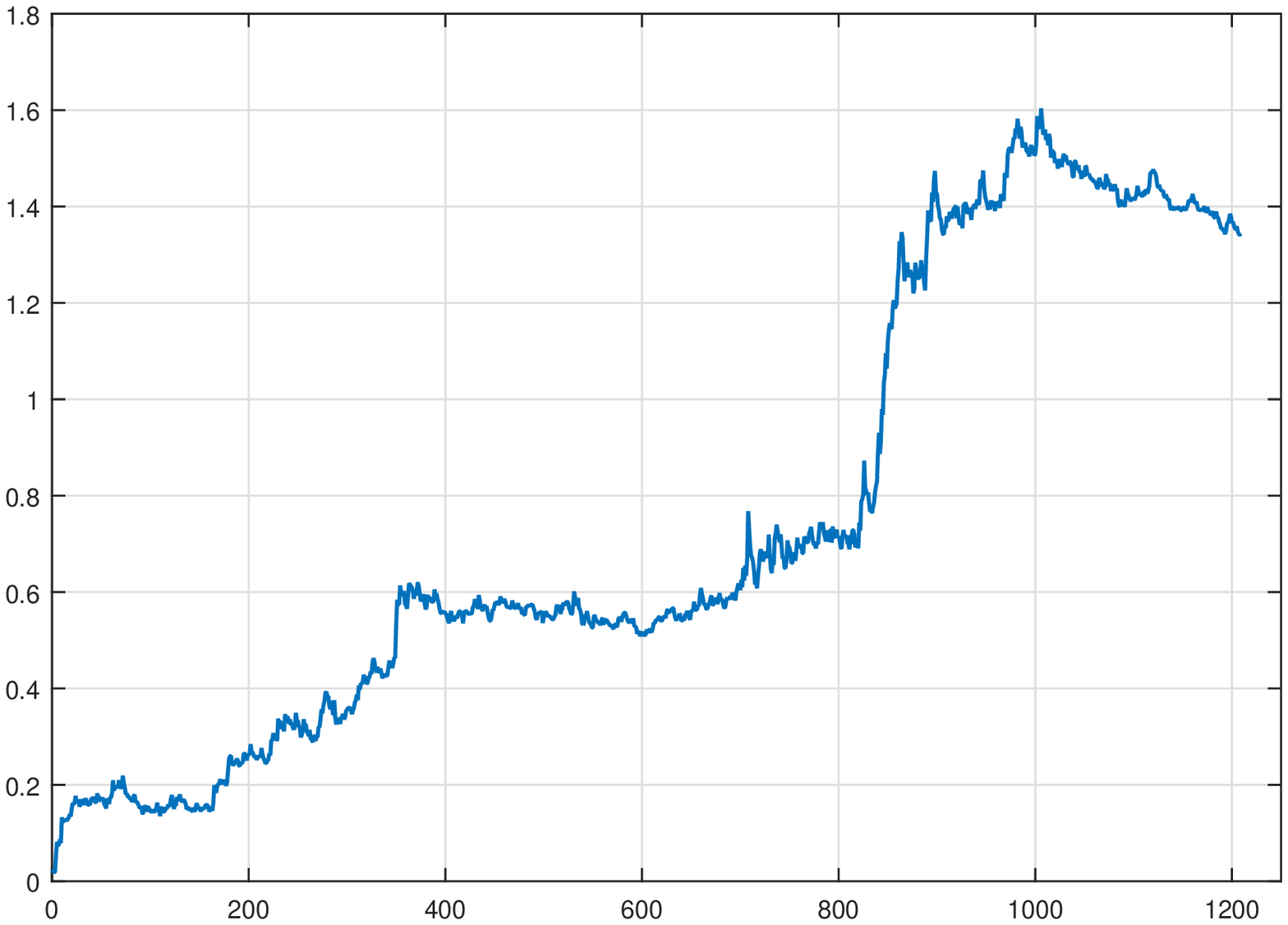}}
\subfigure[]{\includegraphics[width=8cm,height=6cm]{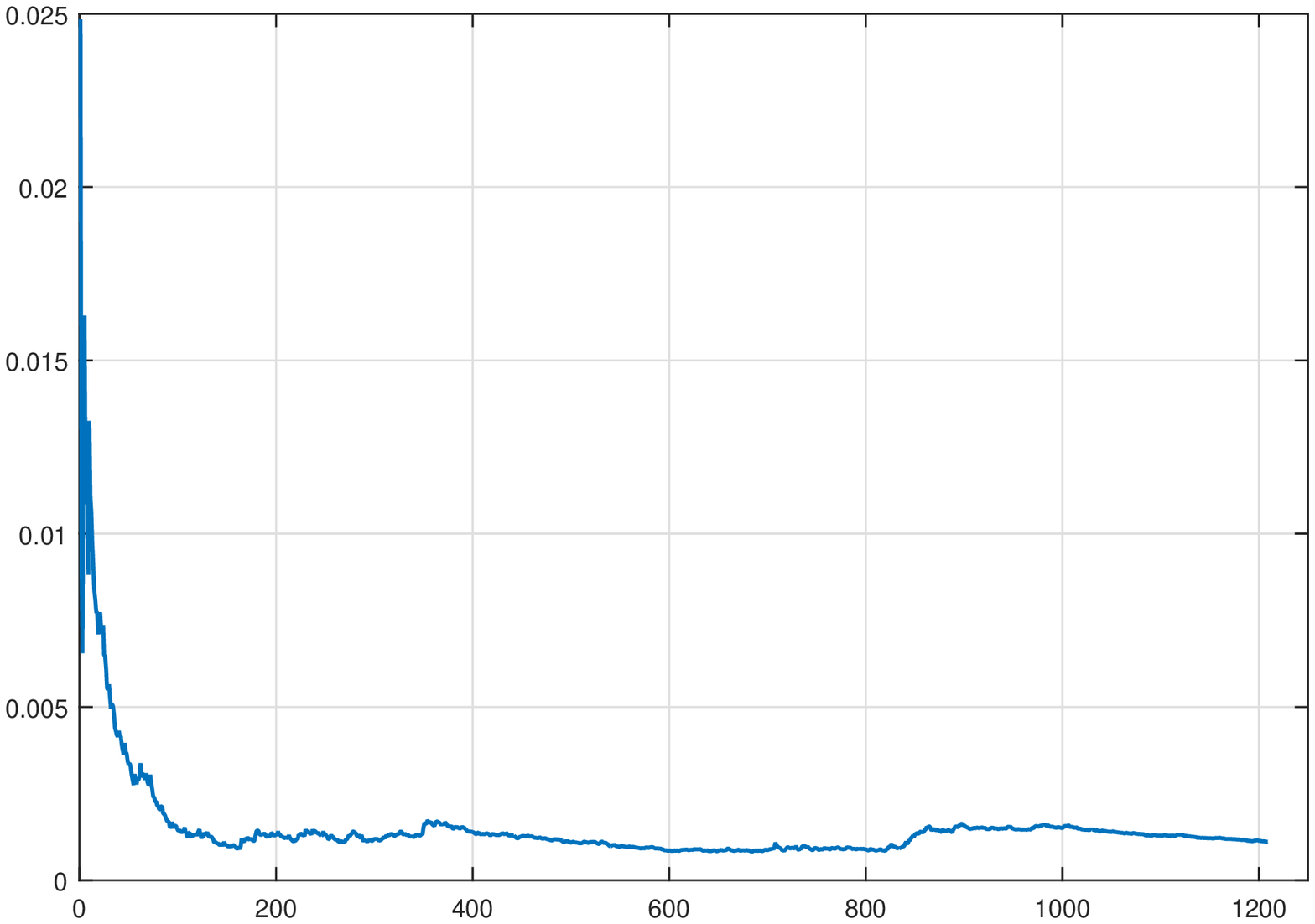}}
\caption{The result of Boll(120,0.1) with 60s data}
\label{FIG4}
\end{figure}

Finally, we check whether the log-return generated by all the trading rules are stationary or not. We test the stationary of single log-return of all 279 strategies and all ADF-tests get the result "h = 1", i.e. the log-return is stationary. Fig. \ref{FIG4} shows the result of $Boll(120,0.1)$ with 60s data. From Fig. \ref{FIG4} (a), the curve follows a trend of rising. Fig. \ref{FIG4} (b) is the average of the cumulative logarithmic return. At the very beginning, the trend is one of violent shaking; however, over time, this trend begins to calm down and converges to a determined value above $5\times10^{-5}$.


\subsection{Data snooping}

Table \ref{TAB5} presents number of significant strategies among all the 297 strategies for each high-frequency data. 54 strategies under the significant level of 5\% and 70 strategies under the significant level of 10\% are significant. Some profitable strategies do not pass SPA test. Two reasons may explain this phenomenon, the first one is that there may be some strategies succeed occasionally, i.e. the phenomenon of data snooping; the other one is that although SPA test greatly improves the power of test, there still exists chance to neglect methods which have predictive power(the type II error), as mentioned by Hansen \cite{Hansen}. However, strategies that are significant according to SPA test match almost best performed strategies.

Results of Step-SPA test indicate that after correcting for data snooping bias, all stationary indicators have significant predictive power. Although some combination of parameters and data frequency may win by chance, most profitable strategies maintain significant predictive ability. Among the 3 stationary indicators, MA and Bollinger bands perform well with some parameters regardless of the frequency of data. However, KDJ is no longer significant for 15s data under Step-SPA test under significance level of 5\% and 10\%. When transaction is taken into account, no strategy is significant and the reason is the same as mentioned in section \ref{Performance}.

\begin{table}[H]
  \centering
  \begin{threeparttable}
  \caption{Step-SPA test results\label{TAB5}}
    \begin{tabular}{crrrrrr}
    \toprule
    \multirow{2}{*}{Method}&
    \multicolumn{2}{c}{15s}&\multicolumn{2}{c}{30s}&\multicolumn{2}{c}{60s}\cr
    \cmidrule(lr){2-3} \cmidrule(lr){4-5} \cmidrule(lr){6-7}
    &$\alpha = 0.05$ & $\alpha = 0.1$ &$\alpha = 0.05$ & $\alpha = 0.1$ &$\alpha = 0.05$ & $\alpha = 0.1$ \cr
    \midrule
    MA   & 15  & 19  & 22  & 25  & 7  & 14 \cr
    KDJ  & 0   & 0   & 2   & 2   & 3  & 4  \cr
    Boll & 1   & 1   & 3   & 3   & 1  & 2  \cr\hline
    Total& 16  & 20  & 27  & 30  & 11 & 20\cr
    \bottomrule
    \end{tabular}
  \end{threeparttable}

\end{table}

\subsection{Strategy improvement}

As above mentioned, we discuss the profitability of simple stationary technical trading rules with high-frequency data of Chinese Index Futures. In practical trading, however, investors concentrate on both profitability and stability. The first property insures the profit, but there's no perfect strategies in the world that win all the time. A big loss may let former accumulated profit be wiped out in a day. From this perspective, stability may be more significant.

Market changes, no strategy profits once for all. That's the reason why traders in reality debug their strategies all the time. In this section, we choose the best strategy from a strategy pool once in several days, and put the chosen one in the market for a while. The interval to compare performance of strategies in strategy pool is called train period, and define the interval in which the chosen strategy be put in market to be test period. The strategy pool consists of "good" strategies. The criterion "good" stands for a sharp ratio larger than 1.5. The specific strategy name and corresponding index are listed in Table \ref{TAB6}. All strategies in the pool are significant according the result of SPA test. And in this section, transaction cost will be neglected.

The beginning string of each strategy name means the technical indicator the strategy used, the number after '\_' stands for the frequency of data, and the numbers in the brackets are the parameters of the strategy. For example, Boll\_30(120,0.1) means the strategy generated by stationary Bollinger band and 30s data with parameter 120 and 0.1.


\begin{table}[htbp]\scalebox{0.85}{
  \centering
  \begin{threeparttable}
  \caption{Strategy Pool}\label{TAB6}
    \begin{tabular}{lrrrr||lrrrr}
    \toprule
    Startegy & \multicolumn{1}{l}{AR\tnote{1}} & \multicolumn{1}{l}{MDP\tnote{1}} & \multicolumn{1}{l}{AR/MDP\tnote{2}} & \multicolumn{1}{l}{SR\tnote{2}} & Startegy & \multicolumn{1}{l}{AR\tnote{1}} & \multicolumn{1}{l}{MDP\tnote{1}} & \multicolumn{1}{l}{AR/MDP\tnote{2}} & \multicolumn{1}{l}{SR\tnote{2}} \\
    \midrule
    Boll\_30(120,0.1) & 59.31\% & 6.64\% & 8.93  & 2.33  & MA\_30(1,30,0.0001) & 30.74\% & 9.44\% & 3.26  & 1.51 \\
    Boll\_30(120,0.5) & 37.61\% & 9.69\% & 3.88  & 1.63  & MA\_30(1,60,0.0001) & 38.47\% & 8.31\% & 4.63  & 1.92 \\
    Boll\_60(60,0.1) & 44.55\% & 9.38\% & 4.75  & 1.79  & MA\_30(1,60,0.0005) & 37.63\% & 12.28\% & 3.06  & 1.68 \\
    KDJR\_30(14,3,3) & 28.86\% & 11.76\% & 2.45  & 1.99  & MA\_30(5,60,0.0001) & 30.83\% & 6.10\% & 5.05  & 1.64 \\
    KDJR\_60(5,1,3) & 33.90\% & 11.68\% & 2.90  & 2.24  & MA\_30(5,60,0.0005) & 39.89\% & 6.45\% & 6.19  & 1.86 \\
    KDJR\_60(9,3,3) & 29.08\% & 8.40\% & 3.46  & 2.10  & MA\_30(5,60,0.001) & 33.70\% & 6.14\% & 5.49  & 1.51 \\
    MA\_15(10,30,0.0001) & 35.96\% & 10.48\% & 3.43  & 1.74  & MA\_30(1,120,0.0001) & 42.61\% & 6.61\% & 6.45  & 2.09 \\
    MA\_15(5,60,0.0001) & 40.36\% & 7.76\% & 5.20  & 1.90  & MA\_30(1,120,0.0005) & 40.84\% & 7.12\% & 5.74  & 1.84 \\
    MA\_15(10,60,0.0005) & 31.80\% & 7.99\% & 3.98  & 1.50  & MA\_30(1,120,0.001) & 38.20\% & 8.96\% & 4.26  & 1.67 \\
    MA\_15(15,60,0.0001) & 39.46\% & 9.79\% & 4.03  & 1.86  & MA\_30(1,120,0.0015) & 36.60\% & 9.66\% & 3.79  & 1.57 \\
    MA\_15(5,120,0.0001) & 42.61\% & 6.77\% & 6.29  & 2.09  & MA\_30(5,120,0.0005) & 36.27\% & 10.03\% & 3.62  & 1.61 \\
    MA\_15(5,120,0.0005) & 45.67\% & 4.54\% & 10.05 & 1.94  & MA\_30(5,120,0.001) & 41.30\% & 9.58\% & 4.31  & 1.83 \\
    MA\_15(5,120,0.001) & 37.71\% & 4.63\% & 8.15  & 1.59  & MA\_30(5,120,0.0015) & 35.19\% & 10.18\% & 3.46  & 1.62 \\
    MA\_15(10,120,0.0001) & 48.84\% & 6.33\% & 7.71  & 2.17  & MA\_30(10,120,0.001) & 35.79\% & 10.21\% & 3.51  & 1.56 \\
    MA\_15(10,120,0.0005) & 45.25\% & 5.71\% & 7.92  & 2.00  & MA\_30(10,120,0.0015) & 35.19\% & 9.66\% & 3.64  & 1.57 \\
    MA\_15(10,120,0.001) & 34.84\% & 6.07\% & 5.74  & 1.56  & MA\_30(15,120,0.001) & 36.44\% & 9.51\% & 3.83  & 1.55 \\
    MA\_15(15,120,0.0001) & 35.18\% & 7.23\% & 4.87  & 1.70  & MA\_60(1,60,0.0005) & 37.87\% & 7.76\% & 4.88  & 1.70 \\
    MA\_15(15,120,0.0005) & 34.81\% & 7.56\% & 4.61  & 1.52  & MA\_60(15,60,0.001) & 36.75\% & 11.15\% & 3.30  & 1.58 \\
    \bottomrule
    \end{tabular}%
    {\tiny
    \begin{tablenotes}
    \footnotesize
    \item[1] AR stands for annual return, MDP means maximum drawback percentage.
    \item[2] AR/MDP is the ratio of AR over MDP and SR is sharp ratio index.
    \end{tablenotes}}
  \end{threeparttable}}
\end{table}%

As for the length of train period and test period, we set test period from 10 days to 80 days every 10 days and train period from 20 days to 80 days every 10 days, and demand that test period is not longer than train period. So we get 35 group train-test period parameters. To make it clear, here explain the operation with the first group parameter. From day 1 to day 20, all strategies in strategy pool run together and we pick the best one, named strategy A. Put strategy A in the market from day 21 to day 30. In a while, all strategies run together from day 11 to day 30 in order to generate the next best strategy B, and then let strategy B run in the market form day 31 to 40. The rest process can be done in the same manner. Strategy A and B can be the same strategy if this strategy perform best both in train period 1 and 2. Strategy with largest Annual return/Max drawback percentage will be the best strategy in a train period.

\begin{table}[htbp]
  \centering
  \begin{threeparttable}
  \caption{Performance of Optimized Strategies}\label{TAB7}
    \begin{tabular}{rrrrrr||rrrrrr}
    \toprule
    \multicolumn{1}{l}{Train\tnote{1}} & \multicolumn{1}{l}{Test\tnote{1}} & \multicolumn{1}{l}{AR\tnote{2}} & \multicolumn{1}{l}{MDP\tnote{2}} & \multicolumn{1}{l}{AR/MDP\tnote{3}} & \multicolumn{1}{l}{SR\tnote{3}} & \multicolumn{1}{l}{Train\tnote{1}} & \multicolumn{1}{l}{Test\tnote{1}} & \multicolumn{1}{l}{AR\tnote{2}} & \multicolumn{1}{l}{MDP\tnote{2}} & \multicolumn{1}{l}{AR/MDP\tnote{3}} & \multicolumn{1}{l}{SR\tnote{3}} \\
    \midrule
    20    & 10    & 47.68\% & 6.96\% & 6.85  & 2.35  & 60    & 50    & 50.46\% & 8.28\% & 6.09  & 2.30 \\
    20    & 20    & 44.44\% & 5.14\% & 8.65  & 2.19  & 60    & 60    & 36.96\% & 9.37\% & 3.94  & 1.71 \\
    30    & 10    & 48.54\% & 9.17\% & 5.29  & 2.30  & 70    & 10    & 42.21\% & 10.76\% & 3.92  & 1.93 \\
    30    & 20    & 49.74\% & 5.39\% & 9.22  & 2.46  & 70    & 20    & 40.78\% & 7.50\% & 5.44  & 1.87 \\
    30    & 30    & 47.57\% & 6.44\% & 7.39  & 2.30  & 70    & 30    & 50.08\% & 5.13\% & 9.76  & 2.41 \\
    40    & 10    & 42.28\% & 8.44\% & 5.01  & 2.08  & 70    & 40    & 46.02\% & 9.69\% & 4.75  & 2.22 \\
    40    & 20    & 47.26\% & 3.33\% & 14.18 & 2.31  & 70    & 50    & 50.80\% & 5.81\% & 8.75  & 2.42 \\
    40    & 30    & 46.99\% & 6.64\% & 7.08  & 2.35  & 70    & 60    & 45.60\% & 9.16\% & 4.98  & 2.11 \\
    40    & 40    & 40.25\% & 7.34\% & 5.48  & 1.93  & 70    & 70    & 53.62\% & 9.41\% & 5.70  & 2.36 \\
    50    & 10    & 53.58\% & 3.53\% & 15.16 & 2.34  & 80    & 10    & 50.49\% & 11.38\% & 4.44  & 2.26 \\
    50    & 20    & 46.40\% & 4.86\% & 9.55  & 2.18  & 80    & 20    & 53.16\% & 5.49\% & 9.68  & 2.37 \\
    50    & 30    & 50.54\% & 4.28\% & 11.80 & 2.43  & 80    & 30    & 51.73\% & 6.01\% & 8.61  & 2.35 \\
    50    & 40    & 38.97\% & 6.42\% & 6.07  & 1.82  & 80    & 40    & 50.74\% & 10.02\% & 5.06  & 2.21 \\
    50    & 50    & 38.54\% & 4.38\% & 8.79  & 1.71  & 80    & 50    & 53.65\% & 7.29\% & 7.36  & 2.40 \\
    60    & 10    & 44.96\% & 10.56\% & 4.26  & 2.01  & 80    & 60    & 47.48\% & 9.04\% & 5.25  & 2.22 \\
    60    & 20    & 39.57\% & 5.82\% & 6.80  & 1.85  & 80    & 70    & 48.29\% & 10.17\% & 4.75  & 2.21 \\
    60    & 30    & 40.35\% & 8.07\% & 5.00  & 1.82  & 80    & 80    & 38.22\% & 10.56\% & 3.62  & 1.79 \\
    60    & 40    & 46.97\% & 6.21\% & 7.57  & 2.25  &       &       &       &       &       &  \\
    \bottomrule
    \end{tabular}%
    {\tiny
    \begin{tablenotes}
    \footnotesize
    \item[1] Train represent for the length of train period while Test means the length of test period.
    \item[2] AR stands for annual return, MDP means maximum drawback percentage.
    \item[3] AR/MDP is the ratio of AR over MDP and SR is sharp ratio index.
    \end{tablenotes}}
  \end{threeparttable}
\end{table}%

\begin{figure}[H]
\centering
\subfigure[]{\includegraphics[width=8cm,height=6cm]{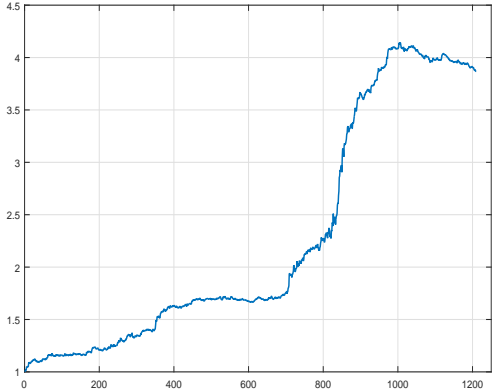}}
\subfigure[]{\includegraphics[width=8cm,height=6cm]{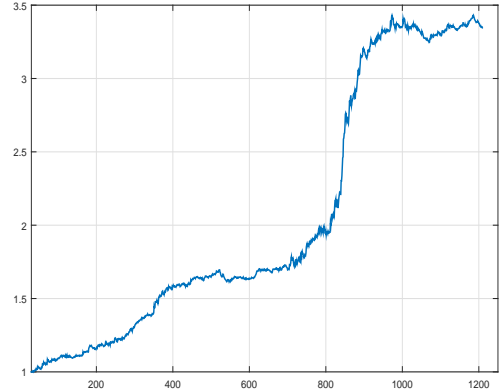}}
\caption{Net Value Curve of Original and Optimized Strategies}
\label{FIG5}
\end{figure}

Table \ref{TAB7} lists the performance of optimized strategies with different length of train period and test period. From the point of sharp ratio, the best strategy in strategy pool is Boll\_30(120,0.1) whose sharp ratio is 2.33 and its ratio of annual return and max drawback percent is 8.93. The highest sharp ratio of optimized strategies is 2.46 with parameter (30,20). The ratio of annual return and max drawback percent is 9.22, which is larger than 8.93. Fig. \ref{FIG5}(a) is the net value curve of Boll\_30(120,0.1) and Fig. \ref{FIG5}(b) indicates the net value curve of the optimized strategy with parameter (30,20). Above we can say the optimized strategy is more stable than single stationary technical indicator strategy. This may be a feasible way to further study.

\section{Conclusion\label{Section5}}

In this paper, we check the profitability of simple stationary technical trading rules with three high-frequency data(15s,30s,60s) of Chinese Index Futures by means of several performance and risk measure and two statistical tests, ADF-test and SPA test. The trading rules consist of stationary MA, stationary Bollinger bands and KDJ rules. Logarithmic return of all 297 strategies passes the ADF-tests regardless of the number of lagged difference terms ($lags$) that are taken. For each stationary technical indicator, there exists significant combination of parameters according to SPA test. And strategies which perform excellently according to performance measure and risk measure all pass SPA test when there are no transaction costs. When transaction costs are taken into account, the best strategy no longer has superiority over others and even loss due to high transaction after September 2nd 2015. At the end of this paper, we propose a feasible method to improve the stability of our strategies from the risk perspective.

\section{Acknowledgments}

We thank Professor Weian Zheng for theoretical direction and helpful discussions, which accelerated our research.

\begin{appendices}
\begin{table}[htbp]\scalebox{0.775}{
  \centering
  \caption{Performance of stationary MA strategy with 15s data}
    \begin{tabular}{rrrrrrrrrrrrrr}
    \toprule
    \multicolumn{1}{c}{$n_s$} & \multicolumn{1}{c}{$n_l$} & \multicolumn{1}{c}{$b$} & \multicolumn{1}{c}{LTN} & \multicolumn{1}{c}{STN} & \multicolumn{1}{c}{ASP} & \multicolumn{1}{c}{ADP} & \multicolumn{1}{c}{AR} & \multicolumn{1}{c}{MDP} & \multicolumn{1}{c}{AR/MDP} & \multicolumn{1}{c}{SR} & \multicolumn{1}{c}{PnL} & \multicolumn{1}{c}{WR} & \multicolumn{1}{c}{AP/AL} \\
    \midrule
    1     & 20    & 0.0001 & 60740 & 62716 & -6.31 & -644.07 & -16.10\% & 85.93\% & -0.19 & -0.84 & -0.15 & 40.03\% & 1.27 \\
    1     & 20    & 0.0005 & 56679 & 58456 & -11.45 & -1090.62 & -27.27\% & 131.70\% & -0.21 & -1.43 & -0.25 & 33.91\% & 1.46 \\
    1     & 20    & 0.001 & 31471 & 32290 & -9.83 & -518.61 & -12.97\% & 80.67\% & -0.16 & -0.73 & -0.16 & 34.33\% & 1.60 \\
    1     & 20    & 0.0015 & 18026 & 18102 & -1.46 & -43.62 & -1.09\% & 45.43\% & -0.02 & -0.07 & -0.02 & 34.66\% & 1.77 \\
    5     & 20    & 0.0001 & 37345 & 38101 & -2.84 & -177.17 & -4.43\% & 46.10\% & -0.10 & -0.20 & -0.04 & 43.67\% & 1.23 \\
    5     & 20    & 0.0005 & 24656 & 25010 & 9.58  & 393.50 & 9.84\% & 16.28\% & 0.60  & 0.49  & 0.11  & 45.91\% & 1.31 \\
    5     & 20    & 0.001 & 11972 & 11688 & 37.81 & 739.90 & 18.50\% & 9.96\% & 1.86  & 1.10  & 0.26  & 44.09\% & 1.63 \\
    5     & 20    & 0.0015 & 6344  & 6005  & 70.88 & 723.97 & 18.10\% & 6.44\% & 2.81  & 1.30  & 0.34  & 38.79\% & 1.89 \\
    10    & 20    & 0.0001 & 32689 & 33126 & 0.27  & 14.84 & 0.37\% & 36.20\% & 0.01  & 0.02  & 0.00  & 48.80\% & 1.05 \\
    10    & 20    & 0.0005 & 14873 & 14723 & 18.74 & 458.71 & 11.47\% & 11.36\% & 1.01  & 0.67  & 0.15  & 47.97\% & 1.26 \\
    10    & 20    & 0.001 & 5548  & 5158  & 45.15 & 399.85 & 10.00\% & 9.56\% & 1.05  & 0.69  & 0.20  & 40.36\% & 1.41 \\
    10    & 20    & 0.0015 & 2512  & 2363  & 68.69 & 276.97 & 6.92\% & 17.66\% & 0.39  & 0.56  & 0.21  & 24.90\% & 1.77 \\
    15    & 20    & 0.0001 & 32404 & 32646 & 6.49  & 349.18 & 8.73\% & 26.26\% & 0.33  & 0.44  & 0.09  & 48.88\% & 1.14 \\
    15    & 20    & 0.0005 & 6257  & 5811  & -23.34 & -232.95 & -5.82\% & 35.22\% & -0.17 & -0.40 & -0.13 & 35.90\% & 1.19 \\
    15    & 20    & 0.001 & 1444  & 1401  & 37.64 & 88.59 & 2.21\% & 23.68\% & 0.09  & 0.19  & 0.10  & 14.72\% & 1.85 \\
    15    & 20    & 0.0015 & 506   & 555   & 68.48 & 60.10 & 1.50\% & 10.14\% & 0.15  & 0.20  & 0.12  & 7.61\% & 1.52 \\
    1     & 30    & 0.0001 & 48983 & 50545 & 4.34  & 356.87 & 8.92\% & 37.26\% & 0.24  & 0.51  & 0.09  & 44.33\% & 1.37 \\
    1     & 30    & 0.0005 & 51587 & 52820 & -6.56 & -566.35 & -14.16\% & 83.92\% & -0.17 & -0.74 & -0.14 & 36.72\% & 1.48 \\
    1     & 30    & 0.001 & 32391 & 33761 & -8.31 & -454.54 & -11.36\% & 76.92\% & -0.15 & -0.61 & -0.13 & 35.98\% & 1.54 \\
    1     & 30    & 0.0015 & 20241 & 20677 & -5.96 & -201.74 & -5.04\% & 62.51\% & -0.08 & -0.29 & -0.07 & 33.33\% & 1.80 \\
    5     & 30    & 0.0001 & 28814 & 29070 & 21.74 & 1040.74 & 26.02\% & 14.72\% & 1.77  & 1.19  & 0.23  & 48.06\% & 1.39 \\
    5     & 30    & 0.0005 & 23200 & 23515 & 12.18 & 470.62 & 11.77\% & 16.90\% & 0.70  & 0.59  & 0.12  & 45.82\% & 1.33 \\
    5     & 30    & 0.001 & 13460 & 13539 & 19.30 & 430.92 & 10.77\% & 14.07\% & 0.77  & 0.62  & 0.14  & 44.25\% & 1.43 \\
    5     & 30    & 0.0015 & 7998  & 7888  & 45.83 & 602.13 & 15.05\% & 10.50\% & 1.43  & 0.94  & 0.23  & 40.12\% & 1.73 \\
    10    & 30    & 0.0001 & 23973 & 24124 & 36.15 & 1438.26 & 35.96\% & 10.48\% & 3.43  & 1.74  & 0.31  & 52.85\% & 1.28 \\
    10    & 30    & 0.0005 & 16367 & 16437 & 36.98 & 1003.33 & 25.08\% & 11.63\% & 2.16  & 1.35  & 0.27  & 48.39\% & 1.44 \\
    10    & 30    & 0.001 & 8329  & 8286  & 34.08 & 468.39 & 11.71\% & 15.83\% & 0.74  & 0.69  & 0.18  & 41.36\% & 1.58 \\
    10    & 30    & 0.0015 & 4554  & 4370  & -5.46 & -40.30 & -1.01\% & 31.80\% & -0.03 & -0.07 & -0.02 & 33.91\% & 1.48 \\
    15    & 30    & 0.0001 & 22515 & 22741 & 20.09 & 751.91 & 18.80\% & 24.17\% & 0.78  & 0.88  & 0.17  & 50.29\% & 1.19 \\
    15    & 30    & 0.0005 & 12381 & 12359 & 32.92 & 673.70 & 16.84\% & 15.82\% & 1.06  & 0.85  & 0.21  & 47.39\% & 1.37 \\
    15    & 30    & 0.001 & 5259  & 5067  & -12.10 & -103.37 & -2.58\% & 36.28\% & -0.07 & -0.15 & -0.05 & 35.81\% & 1.39 \\
    15    & 30    & 0.0015 & 2577  & 2417  & -74.99 & -309.78 & -7.74\% & 47.17\% & -0.16 & -0.55 & -0.19 & 24.48\% & 1.41 \\
    1     & 60    & 0.0001 & 34413 & 35137 & 13.16 & 757.12 & 18.93\% & 24.08\% & 0.79  & 0.98  & 0.18  & 44.50\% & 1.51 \\
    1     & 60    & 0.0005 & 40470 & 41085 & 11.82 & 797.52 & 19.94\% & 33.08\% & 0.60  & 0.89  & 0.17  & 42.68\% & 1.61 \\
    1     & 60    & 0.001 & 30803 & 31516 & 6.65  & 342.53 & 8.56\% & 51.19\% & 0.17  & 0.39  & 0.08  & 39.21\% & 1.69 \\
    1     & 60    & 0.0015 & 21863 & 22469 & 2.75  & 100.69 & 2.52\% & 48.73\% & 0.05  & 0.13  & 0.03  & 35.48\% & 1.85 \\
    5     & 60    & 0.0001 & 19160 & 19373 & 50.65 & 1614.34 & 40.36\% & 7.76\% & 5.20  & 1.90  & 0.33  & 50.54\% & 1.45 \\
    5     & 60    & 0.0005 & 18364 & 18528 & 39.72 & 1211.96 & 30.30\% & 8.62\% & 3.51  & 1.30  & 0.26  & 47.81\% & 1.46 \\
    5     & 60    & 0.001 & 13246 & 13562 & 33.68 & 746.80 & 18.67\% & 15.82\% & 1.18  & 0.87  & 0.19  & 45.16\% & 1.49 \\
    5     & 60    & 0.0015 & 9400  & 9452  & 28.70 & 447.54 & 11.19\% & 13.94\% & 0.80  & 0.58  & 0.14  & 41.27\% & 1.61 \\
    10    & 60    & 0.0001 & 15032 & 15176 & 49.30 & 1231.71 & 30.79\% & 12.21\% & 2.52  & 1.43  & 0.25  & 51.20\% & 1.27 \\
    10    & 60    & 0.0005 & 13276 & 13481 & 57.48 & 1272.16 & 31.80\% & 7.99\% & 3.98  & 1.50  & 0.28  & 48.72\% & 1.45 \\
    10    & 60    & 0.001 & 9379  & 9439  & 38.10 & 593.00 & 14.83\% & 11.42\% & 1.30  & 0.75  & 0.16  & 44.33\% & 1.49 \\
    10    & 60    & 0.0015 & 6409  & 6484  & 24.15 & 257.57 & 6.44\% & 14.32\% & 0.45  & 0.36  & 0.09  & 40.12\% & 1.56 \\
    15    & 60    & 0.0001 & 13352 & 13395 & 71.35 & 1578.46 & 39.46\% & 9.79\% & 4.03  & 1.86  & 0.32  & 51.53\% & 1.37 \\
    15    & 60    & 0.0005 & 11207 & 11302 & 46.17 & 859.60 & 21.49\% & 10.39\% & 2.07  & 1.02  & 0.20  & 47.97\% & 1.35 \\
    15    & 60    & 0.001 & 7571  & 7586  & 13.59 & 170.37 & 4.26\% & 17.39\% & 0.24  & 0.23  & 0.05  & 41.94\% & 1.43 \\
    15    & 60    & 0.0015 & 4976  & 4973  & -22.01 & -181.09 & -4.53\% & 31.47\% & -0.14 & -0.27 & -0.07 & 36.56\% & 1.46 \\
    1     & 120   & 0.0001 & 24723 & 24540 & 27.60 & 1124.76 & 28.12\% & 17.70\% & 1.59  & 1.39  & 0.26  & 45.74\% & 1.59 \\
    1     & 120   & 0.0005 & 30545 & 31296 & 24.13 & 1234.14 & 30.85\% & 20.81\% & 1.48  & 1.31  & 0.24  & 45.82\% & 1.56 \\
    1     & 120   & 0.001 & 25833 & 26998 & 24.96 & 1090.87 & 27.27\% & 24.38\% & 1.12  & 1.14  & 0.22  & 42.10\% & 1.76 \\
    1     & 120   & 0.0015 & 20999 & 21336 & 28.36 & 993.20 & 24.83\% & 31.55\% & 0.79  & 1.05  & 0.22  & 40.03\% & 1.90 \\
    5     & 120   & 0.0001 & 13465 & 13399 & 76.70 & 1704.37 & 42.61\% & 6.77\% & 6.29  & 2.09  & 0.35  & 51.28\% & 1.44 \\
    5     & 120   & 0.0005 & 13575 & 13819 & 80.61 & 1826.60 & 45.67\% & 4.54\% & 10.05 & 1.94  & 0.35  & 50.04\% & 1.53 \\
    5     & 120   & 0.001 & 11340 & 11598 & 79.51 & 1508.54 & 37.71\% & 4.63\% & 8.15  & 1.59  & 0.31  & 46.48\% & 1.66 \\
    5     & 120   & 0.0015 & 8976  & 9128  & 86.30 & 1292.26 & 32.31\% & 6.43\% & 5.02  & 1.47  & 0.30  & 44.58\% & 1.74 \\
    10    & 120   & 0.0001 & 10353 & 10384 & 113.90 & 1953.70 & 48.84\% & 6.33\% & 7.71  & 2.17  & 0.36  & 51.94\% & 1.45 \\
    10    & 120   & 0.0005 & 9750  & 9859  & 111.61 & 1810.17 & 45.25\% & 5.71\% & 7.92  & 2.00  & 0.35  & 51.03\% & 1.46 \\
    10    & 120   & 0.001 & 8067  & 8240  & 103.31 & 1393.45 & 34.84\% & 6.07\% & 5.74  & 1.56  & 0.30  & 46.57\% & 1.65 \\
    10    & 120   & 0.0015 & 6385  & 6422  & 111.64 & 1182.58 & 29.56\% & 8.59\% & 3.44  & 1.35  & 0.28  & 44.58\% & 1.68 \\
    15    & 120   & 0.0001 & 8928  & 8951  & 95.15 & 1407.05 & 35.18\% & 7.23\% & 4.87  & 1.70  & 0.28  & 50.29\% & 1.36 \\
    15    & 120   & 0.0005 & 8146  & 8309  & 102.31 & 1392.51 & 34.81\% & 7.56\% & 4.61  & 1.52  & 0.28  & 48.14\% & 1.49 \\
    15    & 120   & 0.001 & 6678  & 6799  & 89.99 & 1003.13 & 25.08\% & 9.39\% & 2.67  & 1.09  & 0.22  & 45.49\% & 1.53 \\
    15    & 120   & 0.0015 & 5171  & 5272  & 114.75 & 991.22 & 24.78\% & 9.57\% & 2.59  & 1.15  & 0.24  & 43.92\% & 1.65 \\
    \bottomrule
    \end{tabular}%
\label{TAB:MA1}}
\end{table}%

\begin{table}[htbp]\scalebox{0.775}{
  \centering
  \caption{Performance of stationary MA strategy with 30s data}
    \begin{tabular}{rrrrrrrrrrrrrr}
    \toprule
    \multicolumn{1}{c}{$n_s$} & \multicolumn{1}{c}{$n_l$} & \multicolumn{1}{c}{$b$} & \multicolumn{1}{c}{LTN} & \multicolumn{1}{c}{STN} & \multicolumn{1}{c}{ASP} & \multicolumn{1}{c}{ADP} & \multicolumn{1}{c}{AR} & \multicolumn{1}{c}{MDP} & \multicolumn{1}{c}{AR/MDP} & \multicolumn{1}{c}{SR} & \multicolumn{1}{c}{PnL} & \multicolumn{1}{c}{WR} & \multicolumn{1}{c}{AP/AL} \\
    \midrule
    1     & 20    & 0.0001 & 27512 & 27918 & 6.61  & 303.13 & 7.58\% & 28.66\% & 0.26  & 0.43  & 0.08  & 43.01\% & 1.44 \\
    1     & 20    & 0.0005 & 31842 & 32553 & 7.62  & 405.71 & 10.14\% & 31.49\% & 0.32  & 0.51  & 0.10  & 43.18\% & 1.45 \\
    1     & 20    & 0.001 & 22224 & 22868 & -0.41 & -15.19 & -0.38\% & 47.66\% & -0.01 & -0.02 & 0.00  & 38.46\% & 1.58 \\
    1     & 20    & 0.0015 & 14541 & 15019 & 6.59  & 161.14 & 4.03\% & 24.24\% & 0.17  & 0.25  & 0.06  & 36.64\% & 1.78 \\
    5     & 20    & 0.0001 & 17749 & 18017 & 28.67 & 848.19 & 21.20\% & 20.78\% & 1.02  & 1.11  & 0.20  & 49.71\% & 1.26 \\
    5     & 20    & 0.0005 & 15328 & 15571 & 34.58 & 883.77 & 22.09\% & 14.74\% & 1.50  & 1.17  & 0.23  & 48.47\% & 1.35 \\
    5     & 20    & 0.001 & 9199  & 9342  & 39.49 & 605.61 & 15.14\% & 8.09\% & 1.87  & 0.91  & 0.20  & 44.75\% & 1.50 \\
    5     & 20    & 0.0015 & 5646  & 5596  & 28.64 & 266.35 & 6.66\% & 18.44\% & 0.36  & 0.43  & 0.11  & 36.97\% & 1.66 \\
    10    & 20    & 0.0001 & 16463 & 16490 & 16.98 & 462.93 & 11.57\% & 31.41\% & 0.37  & 0.58  & 0.11  & 46.65\% & 1.27 \\
    10    & 20    & 0.0005 & 10609 & 10554 & 40.14 & 702.63 & 17.57\% & 15.01\% & 1.17  & 1.01  & 0.22  & 46.15\% & 1.47 \\
    10    & 20    & 0.001 & 4904  & 4820  & -14.07 & -113.15 & -2.83\% & 32.73\% & -0.09 & -0.20 & -0.05 & 37.97\% & 1.32 \\
    10    & 20    & 0.0015 & 2586  & 2381  & -43.00 & -176.67 & -4.42\% & 35.87\% & -0.12 & -0.36 & -0.11 & 27.54\% & 1.38 \\
    15    & 20    & 0.0001 & 17706 & 17601 & 6.93  & 202.38 & 5.06\% & 33.23\% & 0.15  & 0.26  & 0.05  & 45.66\% & 1.25 \\
    15    & 20    & 0.0005 & 5568  & 5431  & -21.97 & -199.90 & -5.00\% & 53.69\% & -0.09 & -0.34 & -0.09 & 39.12\% & 1.19 \\
    15    & 20    & 0.001 & 1571  & 1472  & -70.67 & -177.87 & -4.45\% & 31.60\% & -0.14 & -0.40 & -0.15 & 21.26\% & 1.15 \\
    15    & 20    & 0.0015 & 603   & 621   & -58.97 & -59.70 & -1.49\% & 16.19\% & -0.09 & -0.18 & -0.09 & 10.17\% & 1.26 \\
    1     & 30    & 0.0001 & 22180 & 22934 & 32.96 & 1229.78 & 30.74\% & 9.44\% & 3.26  & 1.51  & 0.28  & 45.00\% & 1.68 \\
    1     & 30    & 0.0005 & 27484 & 28052 & 20.27 & 930.92 & 23.27\% & 21.71\% & 1.07  & 1.14  & 0.21  & 44.42\% & 1.57 \\
    1     & 30    & 0.001 & 21216 & 21779 & 18.18 & 646.50 & 16.16\% & 35.26\% & 0.46  & 0.79  & 0.16  & 39.78\% & 1.79 \\
    1     & 30    & 0.0015 & 15078 & 15495 & 11.15 & 281.89 & 7.05\% & 28.71\% & 0.25  & 0.38  & 0.09  & 36.97\% & 1.85 \\
    5     & 30    & 0.0001 & 13581 & 13693 & 46.37 & 1046.10 & 26.15\% & 14.89\% & 1.76  & 1.30  & 0.24  & 51.94\% & 1.21 \\
    5     & 30    & 0.0005 & 13255 & 13460 & 41.02 & 906.45 & 22.66\% & 11.50\% & 1.97  & 1.12  & 0.21  & 46.90\% & 1.41 \\
    5     & 30    & 0.001 & 9376  & 9465  & 27.15 & 423.08 & 10.58\% & 12.72\% & 0.83  & 0.57  & 0.12  & 44.67\% & 1.40 \\
    5     & 30    & 0.0015 & 6476  & 6530  & 12.47 & 134.19 & 3.35\% & 14.64\% & 0.23  & 0.20  & 0.05  & 39.78\% & 1.50 \\
    10    & 30    & 0.0001 & 11740 & 11809 & 28.76 & 560.15 & 14.00\% & 16.91\% & 0.83  & 0.72  & 0.13  & 50.04\% & 1.15 \\
    10    & 30    & 0.0005 & 9941  & 10023 & 39.03 & 644.42 & 16.11\% & 11.89\% & 1.35  & 0.79  & 0.16  & 45.08\% & 1.44 \\
    10    & 30    & 0.001 & 6248  & 6293  & -3.65 & -37.87 & -0.95\% & 24.94\% & -0.04 & -0.05 & -0.01 & 40.61\% & 1.39 \\
    10    & 30    & 0.0015 & 3941  & 3918  & -56.32 & -366.10 & -9.15\% & 46.75\% & -0.20 & -0.56 & -0.14 & 33.91\% & 1.41 \\
    15    & 30    & 0.0001 & 11405 & 11380 & -18.93 & -356.72 & -8.92\% & 59.59\% & -0.15 & -0.45 & -0.09 & 45.82\% & 1.07 \\
    15    & 30    & 0.0005 & 8174  & 8144  & -8.78 & -118.46 & -2.96\% & 39.39\% & -0.08 & -0.14 & -0.03 & 44.50\% & 1.18 \\
    15    & 30    & 0.001 & 4325  & 4282  & -50.98 & -362.93 & -9.07\% & 47.17\% & -0.19 & -0.50 & -0.13 & 37.14\% & 1.30 \\
    15    & 30    & 0.0015 & 2436  & 2357  & -97.87 & -387.99 & -9.70\% & 50.36\% & -0.19 & -0.63 & -0.19 & 31.18\% & 1.17 \\
    1     & 60    & 0.0001 & 15965 & 15972 & 58.25 & 1538.76 & 38.47\% & 8.31\% & 4.63  & 1.92  & 0.34  & 48.88\% & 1.57 \\
    1     & 60    & 0.0005 & 20654 & 21364 & 43.31 & 1505.36 & 37.63\% & 12.28\% & 3.06  & 1.68  & 0.30  & 45.99\% & 1.68 \\
    1     & 60    & 0.001 & 17976 & 18533 & 45.05 & 1360.35 & 34.01\% & 14.47\% & 2.35  & 1.49  & 0.29  & 43.18\% & 1.83 \\
    1     & 60    & 0.0015 & 14584 & 14730 & 49.74 & 1206.05 & 30.15\% & 18.21\% & 1.66  & 1.33  & 0.27  & 41.11\% & 1.95 \\
    5     & 60    & 0.0001 & 9321  & 9309  & 80.04 & 1233.30 & 30.83\% & 6.10\% & 5.05  & 1.64  & 0.28  & 50.87\% & 1.33 \\
    5     & 60    & 0.0005 & 9664  & 9856  & 98.82 & 1595.58 & 39.89\% & 6.45\% & 6.19  & 1.86  & 0.32  & 51.12\% & 1.40 \\
    5     & 60    & 0.001 & 8112  & 8285  & 99.38 & 1347.89 & 33.70\% & 6.14\% & 5.49  & 1.51  & 0.29  & 45.99\% & 1.65 \\
    5     & 60    & 0.0015 & 6438  & 6521  & 101.63 & 1089.33 & 27.23\% & 8.18\% & 3.33  & 1.25  & 0.26  & 43.42\% & 1.72 \\
    10    & 60    & 0.0001 & 7639  & 7529  & 83.68 & 1049.83 & 26.25\% & 9.50\% & 2.76  & 1.31  & 0.23  & 48.72\% & 1.36 \\
    10    & 60    & 0.0005 & 7282  & 7304  & 103.34 & 1246.70 & 31.17\% & 9.84\% & 3.17  & 1.37  & 0.26  & 46.73\% & 1.53 \\
    10    & 60    & 0.001 & 5841  & 5934  & 97.96 & 954.09 & 23.85\% & 9.84\% & 2.42  & 1.02  & 0.21  & 44.00\% & 1.60 \\
    10    & 60    & 0.0015 & 4452  & 4533  & 108.65 & 807.44 & 20.19\% & 10.73\% & 1.88  & 0.94  & 0.21  & 41.94\% & 1.68 \\
    15    & 60    & 0.0001 & 6756  & 6806  & 56.00 & 628.19 & 15.70\% & 18.98\% & 0.83  & 0.75  & 0.14  & 47.89\% & 1.26 \\
    15    & 60    & 0.0005 & 6148  & 6179  & 71.01 & 724.07 & 18.10\% & 14.36\% & 1.26  & 0.75  & 0.16  & 44.91\% & 1.44 \\
    15    & 60    & 0.001 & 4811  & 4851  & 97.76 & 781.24 & 19.53\% & 12.30\% & 1.59  & 0.82  & 0.18  & 42.93\% & 1.61 \\
    15    & 60    & 0.0015 & 3523  & 3618  & 122.08 & 721.09 & 18.03\% & 13.64\% & 1.32  & 0.81  & 0.19  & 41.52\% & 1.62 \\
    1     & 120   & 0.0001 & 11960 & 12069 & 85.76 & 1704.47 & 42.61\% & 6.61\% & 6.45  & 2.09  & 0.36  & 46.98\% & 1.75 \\
    1     & 120   & 0.0005 & 15843 & 16687 & 60.71 & 1633.55 & 40.84\% & 7.12\% & 5.74  & 1.84  & 0.32  & 48.39\% & 1.56 \\
    1     & 120   & 0.001 & 14409 & 15566 & 61.63 & 1528.04 & 38.20\% & 8.96\% & 4.26  & 1.67  & 0.30  & 45.33\% & 1.72 \\
    1     & 120   & 0.0015 & 12260 & 13190 & 69.56 & 1464.17 & 36.60\% & 9.66\% & 3.79  & 1.57  & 0.30  & 42.51\% & 1.92 \\
    5     & 120   & 0.0001 & 6785  & 7062  & 77.61 & 888.88 & 22.22\% & 9.36\% & 2.37  & 1.19  & 0.21  & 47.06\% & 1.42 \\
    5     & 120   & 0.0005 & 7300  & 7589  & 117.81 & 1450.82 & 36.27\% & 10.03\% & 3.62  & 1.61  & 0.28  & 49.63\% & 1.41 \\
    5     & 120   & 0.001 & 6363  & 6753  & 152.26 & 1651.86 & 41.30\% & 9.58\% & 4.31  & 1.83  & 0.32  & 48.06\% & 1.60 \\
    5     & 120   & 0.0015 & 5387  & 5820  & 151.83 & 1407.44 & 35.19\% & 10.18\% & 3.46  & 1.62  & 0.30  & 45.24\% & 1.72 \\
    10    & 120   & 0.0001 & 5603  & 5548  & 79.71 & 735.24 & 18.38\% & 14.27\% & 1.29  & 0.91  & 0.16  & 46.40\% & 1.37 \\
    10    & 120   & 0.0005 & 5398  & 5512  & 140.91 & 1271.61 & 31.79\% & 12.21\% & 2.60  & 1.36  & 0.25  & 47.23\% & 1.48 \\
    10    & 120   & 0.001 & 4568  & 4811  & 184.56 & 1431.76 & 35.79\% & 10.21\% & 3.51  & 1.56  & 0.29  & 46.73\% & 1.59 \\
    10    & 120   & 0.0015 & 3815  & 4076  & 215.68 & 1407.69 & 35.19\% & 9.66\% & 3.64  & 1.57  & 0.30  & 44.33\% & 1.75 \\
    15    & 120   & 0.0001 & 4837  & 4899  & 146.62 & 1180.69 & 29.52\% & 9.52\% & 3.10  & 1.29  & 0.23  & 47.15\% & 1.46 \\
    15    & 120   & 0.0005 & 4495  & 4623  & 176.92 & 1334.29 & 33.36\% & 13.63\% & 2.45  & 1.35  & 0.25  & 45.99\% & 1.56 \\
    15    & 120   & 0.001 & 3735  & 3961  & 228.96 & 1457.47 & 36.44\% & 9.51\% & 3.83  & 1.55  & 0.29  & 45.74\% & 1.66 \\
    15    & 120   & 0.0015 & 3090  & 3302  & 253.50 & 1340.25 & 33.51\% & 9.38\% & 3.57  & 1.49  & 0.29  & 43.18\% & 1.80 \\
    \bottomrule
    \end{tabular}%
  \label{TAB:MA2}}%
\end{table}%

\begin{table}[htbp]\scalebox{0.775}{
  \centering
  \caption{Performance of stationary MA strategy with 60s data}
    \begin{tabular}{rrrrrrrrrrrrrr}
    \toprule
    \multicolumn{1}{c}{$n_s$} & \multicolumn{1}{c}{$n_l$} & \multicolumn{1}{c}{$b$} & \multicolumn{1}{c}{LTN} & \multicolumn{1}{c}{STN} & \multicolumn{1}{c}{ASP} & \multicolumn{1}{c}{ADP} & \multicolumn{1}{c}{AR} & \multicolumn{1}{c}{MDP} & \multicolumn{1}{c}{AR/MDP} & \multicolumn{1}{c}{SR} & \multicolumn{1}{c}{PnL} & \multicolumn{1}{c}{WR} & \multicolumn{1}{c}{AP/AL} \\
    \midrule
    1     & 20    & 0.0001 & 12664 & 13131 & 19.39 & 413.60 & 10.34\% & 22.92\% & 0.45  & 0.50  & 0.11  & 45.16\% & 1.35 \\
    1     & 20    & 0.0005 & 16816 & 17098 & 25.65 & 719.40 & 17.99\% & 24.42\% & 0.74  & 0.86  & 0.16  & 44.91\% & 1.46 \\
    1     & 20    & 0.001 & 13876 & 14281 & 24.83 & 578.31 & 14.46\% & 21.94\% & 0.66  & 0.67  & 0.14  & 42.76\% & 1.54 \\
    1     & 20    & 0.0015 & 10669 & 10812 & 25.71 & 456.87 & 11.42\% & 21.42\% & 0.53  & 0.56  & 0.13  & 40.36\% & 1.66 \\
    5     & 20    & 0.0001 & 8342  & 8349  & -9.37 & -129.38 & -3.23\% & 42.93\% & -0.08 & -0.16 & -0.03 & 44.09\% & 1.23 \\
    5     & 20    & 0.0005 & 8712  & 8860  & 0.36  & 5.21  & 0.13\% & 32.13\% & 0.00  & 0.01  & 0.00  & 43.84\% & 1.27 \\
    5     & 20    & 0.001 & 6419  & 6425  & -16.55 & -175.83 & -4.40\% & 38.76\% & -0.11 & -0.20 & -0.05 & 41.03\% & 1.35 \\
    5     & 20    & 0.0015 & 4468  & 4490  & -57.12 & -423.23 & -10.58\% & 54.09\% & -0.20 & -0.51 & -0.13 & 36.81\% & 1.38 \\
    10    & 20    & 0.0001 & 8048  & 8012  & -42.96 & -570.62 & -14.27\% & 78.98\% & -0.18 & -0.65 & -0.13 & 43.26\% & 1.13 \\
    10    & 20    & 0.0005 & 6498  & 6649  & -60.09 & -653.40 & -16.33\% & 84.80\% & -0.19 & -0.71 & -0.16 & 42.60\% & 1.13 \\
    10    & 20    & 0.001 & 3817  & 3854  & -14.75 & -93.60 & -2.34\% & 45.42\% & -0.05 & -0.11 & -0.03 & 40.20\% & 1.27 \\
    10    & 20    & 0.0015 & 2305  & 2227  & -28.50 & -106.85 & -2.67\% & 33.28\% & -0.08 & -0.14 & -0.05 & 32.75\% & 1.37 \\
    15    & 20    & 0.0001 & 9276  & 9133  & 9.60  & 146.10 & 3.65\% & 35.99\% & 0.10  & 0.15  & 0.03  & 45.82\% & 1.21 \\
    15    & 20    & 0.0005 & 4365  & 4402  & 59.11 & 428.64 & 10.72\% & 24.39\% & 0.44  & 0.51  & 0.13  & 43.67\% & 1.32 \\
    15    & 20    & 0.001 & 1592  & 1498  & 38.23 & 97.72 & 2.44\% & 18.35\% & 0.13  & 0.15  & 0.05  & 25.64\% & 1.43 \\
    15    & 20    & 0.0015 & 685   & 677   & 140.53 & 158.31 & 3.96\% & 15.29\% & 0.26  & 0.29  & 0.14  & 15.72\% & 1.31 \\
    1     & 30    & 0.0001 & 10601 & 10724 & 51.19 & 902.88 & 22.57\% & 10.64\% & 2.12  & 1.15  & 0.22  & 43.76\% & 1.64 \\
    1     & 30    & 0.0005 & 14050 & 14513 & 34.09 & 805.46 & 20.14\% & 19.95\% & 1.01  & 0.95  & 0.18  & 45.41\% & 1.47 \\
    1     & 30    & 0.001 & 12584 & 12856 & 46.88 & 986.50 & 24.66\% & 16.27\% & 1.52  & 1.14  & 0.22  & 43.26\% & 1.68 \\
    1     & 30    & 0.0015 & 10317 & 10396 & 46.45 & 795.73 & 19.89\% & 21.45\% & 0.93  & 0.91  & 0.19  & 41.19\% & 1.75 \\
    5     & 30    & 0.0001 & 6552  & 6589  & 39.86 & 433.20 & 10.83\% & 15.65\% & 0.69  & 0.56  & 0.11  & 46.73\% & 1.27 \\
    5     & 30    & 0.0005 & 7210  & 7327  & 29.64 & 356.43 & 8.91\% & 22.28\% & 0.40  & 0.41  & 0.09  & 44.83\% & 1.34 \\
    5     & 30    & 0.001 & 5888  & 6015  & 32.59 & 320.89 & 8.02\% & 21.64\% & 0.37  & 0.37  & 0.08  & 42.18\% & 1.48 \\
    5     & 30    & 0.0015 & 4530  & 4607  & 51.84 & 391.81 & 9.80\% & 17.60\% & 0.56  & 0.45  & 0.10  & 39.78\% & 1.62 \\
    10    & 30    & 0.0001 & 5777  & 5757  & 17.50 & 167.00 & 4.17\% & 30.25\% & 0.14  & 0.19  & 0.04  & 45.49\% & 1.24 \\
    10    & 30    & 0.0005 & 5476  & 5549  & 26.60 & 242.58 & 6.06\% & 31.80\% & 0.19  & 0.23  & 0.05  & 43.01\% & 1.39 \\
    10    & 30    & 0.001 & 4076  & 4170  & 78.38 & 534.59 & 13.36\% & 18.62\% & 0.72  & 0.54  & 0.13  & 41.03\% & 1.62 \\
    10    & 30    & 0.0015 & 2889  & 2966  & 135.94 & 658.31 & 16.46\% & 16.93\% & 0.97  & 0.75  & 0.19  & 39.37\% & 1.68 \\
    15    & 30    & 0.0001 & 5580  & 5586  & 44.95 & 415.14 & 10.38\% & 23.73\% & 0.44  & 0.44  & 0.09  & 44.75\% & 1.35 \\
    15    & 30    & 0.0005 & 4654  & 4724  & 94.76 & 735.04 & 18.38\% & 13.84\% & 1.33  & 0.75  & 0.16  & 45.82\% & 1.40 \\
    15    & 30    & 0.001 & 3004  & 3108  & 202.83 & 1025.41 & 25.64\% & 7.35\% & 3.49  & 1.19  & 0.27  & 42.27\% & 1.71 \\
    15    & 30    & 0.0015 & 1916  & 1969  & 341.36 & 1096.92 & 27.42\% & 8.03\% & 3.42  & 1.41  & 0.34  & 37.30\% & 1.82 \\
    1     & 60    & 0.0001 & 7920  & 8112  & 67.31 & 892.51 & 22.31\% & 7.96\% & 2.80  & 1.20  & 0.22  & 44.75\% & 1.58 \\
    1     & 60    & 0.0005 & 10739 & 11244 & 83.32 & 1514.94 & 37.87\% & 7.76\% & 4.88  & 1.70  & 0.30  & 46.24\% & 1.66 \\
    1     & 60    & 0.001 & 9974  & 10812 & 72.76 & 1251.02 & 31.28\% & 8.91\% & 3.51  & 1.39  & 0.25  & 45.49\% & 1.59 \\
    1     & 60    & 0.0015 & 8590  & 9303  & 68.90 & 1019.65 & 25.49\% & 12.91\% & 1.97  & 1.17  & 0.22  & 41.69\% & 1.80 \\
    5     & 60    & 0.0001 & 4850  & 4861  & 59.72 & 479.65 & 11.99\% & 22.27\% & 0.54  & 0.59  & 0.11  & 43.01\% & 1.49 \\
    5     & 60    & 0.0005 & 5391  & 5540  & 70.88 & 640.84 & 16.02\% & 23.30\% & 0.69  & 0.67  & 0.13  & 45.57\% & 1.37 \\
    5     & 60    & 0.001 & 4627  & 4841  & 145.96 & 1143.08 & 28.58\% & 13.35\% & 2.14  & 1.23  & 0.24  & 45.24\% & 1.57 \\
    5     & 60    & 0.0015 & 3848  & 4120  & 177.03 & 1166.75 & 29.17\% & 11.14\% & 2.62  & 1.30  & 0.25  & 43.26\% & 1.72 \\
    10    & 60    & 0.0001 & 3991  & 4041  & 146.22 & 971.41 & 24.29\% & 13.98\% & 1.74  & 1.10  & 0.21  & 45.16\% & 1.53 \\
    10    & 60    & 0.0005 & 3907  & 4029  & 180.45 & 1184.52 & 29.61\% & 12.09\% & 2.45  & 1.23  & 0.24  & 45.08\% & 1.58 \\
    10    & 60    & 0.001 & 3232  & 3467  & 226.99 & 1257.72 & 31.44\% & 12.67\% & 2.48  & 1.29  & 0.26  & 44.83\% & 1.64 \\
    10    & 60    & 0.0015 & 2707  & 2861  & 233.74 & 1076.48 & 26.91\% & 10.88\% & 2.47  & 1.17  & 0.24  & 43.18\% & 1.68 \\
    15    & 60    & 0.0001 & 3510  & 3518  & 213.53 & 1241.29 & 31.03\% & 12.12\% & 2.56  & 1.40  & 0.25  & 46.73\% & 1.52 \\
    15    & 60    & 0.0005 & 3274  & 3403  & 228.41 & 1261.44 & 31.54\% & 11.10\% & 2.84  & 1.39  & 0.26  & 45.82\% & 1.59 \\
    15    & 60    & 0.001 & 2624  & 2830  & 325.89 & 1470.12 & 36.75\% & 11.15\% & 3.30  & 1.58  & 0.31  & 45.16\% & 1.72 \\
    15    & 60    & 0.0015 & 2195  & 2326  & 340.89 & 1274.74 & 31.87\% & 10.89\% & 2.93  & 1.45  & 0.30  & 42.43\% & 1.80 \\
    1     & 120   & 0.0001 & 6692  & 6928  & 59.41 & 669.33 & 16.73\% & 13.96\% & 1.20  & 0.81  & 0.16  & 42.85\% & 1.59 \\
    1     & 120   & 0.0005 & 9056  & 9707  & 34.69 & 538.31 & 13.46\% & 14.70\% & 0.92  & 0.61  & 0.12  & 44.33\% & 1.42 \\
    1     & 120   & 0.001 & 8408  & 9514  & 53.49 & 792.95 & 19.82\% & 10.97\% & 1.81  & 0.87  & 0.16  & 43.59\% & 1.54 \\
    1     & 120   & 0.0015 & 7244  & 8421  & 51.18 & 663.18 & 16.58\% & 13.74\% & 1.21  & 0.72  & 0.14  & 42.93\% & 1.55 \\
    5     & 120   & 0.0001 & 4046  & 4040  & 53.04 & 354.74 & 8.87\% & 20.48\% & 0.43  & 0.41  & 0.08  & 44.67\% & 1.34 \\
    5     & 120   & 0.0005 & 4485  & 4654  & 65.30 & 493.60 & 12.34\% & 26.10\% & 0.47  & 0.52  & 0.10  & 46.24\% & 1.29 \\
    5     & 120   & 0.001 & 3854  & 4199  & 76.32 & 508.39 & 12.71\% & 20.60\% & 0.62  & 0.52  & 0.11  & 45.16\% & 1.36 \\
    5     & 120   & 0.0015 & 3230  & 3692  & 107.28 & 614.19 & 15.35\% & 25.03\% & 0.61  & 0.66  & 0.13  & 43.01\% & 1.50 \\
    10    & 120   & 0.0001 & 3239  & 3312  & 106.87 & 579.06 & 14.48\% & 12.37\% & 1.17  & 0.66  & 0.13  & 45.74\% & 1.35 \\
    10    & 120   & 0.0005 & 3187  & 3330  & 147.45 & 794.79 & 19.87\% & 19.86\% & 1.00  & 0.82  & 0.16  & 45.99\% & 1.38 \\
    10    & 120   & 0.001 & 2686  & 2914  & 202.96 & 940.10 & 23.50\% & 15.00\% & 1.57  & 0.99  & 0.20  & 46.07\% & 1.44 \\
    10    & 120   & 0.0015 & 2282  & 2583  & 207.79 & 836.13 & 20.90\% & 15.86\% & 1.32  & 0.89  & 0.18  & 42.51\% & 1.62 \\
    15    & 120   & 0.0001 & 2825  & 2814  & 194.58 & 907.54 & 22.69\% & 19.46\% & 1.17  & 0.98  & 0.18  & 46.65\% & 1.40 \\
    15    & 120   & 0.0005 & 2574  & 2720  & 228.01 & 998.41 & 24.96\% & 21.46\% & 1.16  & 1.05  & 0.20  & 47.15\% & 1.39 \\
    15    & 120   & 0.001 & 2135  & 2364  & 299.49 & 1114.49 & 27.86\% & 14.40\% & 1.94  & 1.22  & 0.23  & 44.58\% & 1.60 \\
    15    & 120   & 0.0015 & 1825  & 2081  & 328.79 & 1062.23 & 26.56\% & 9.15\% & 2.90  & 1.21  & 0.24  & 43.34\% & 1.65 \\
    \bottomrule
    \end{tabular}%
  \label{TAB:MA3}}%
\end{table}%

\begin{table}[htbp]\scalebox{0.8}{
  \centering
  \caption{Performance of stationary KDJ strategy with 15s data}
    \begin{tabular}{rrrrrrrrrrrrrr}
    \toprule
    \multicolumn{1}{c}{m} & \multicolumn{1}{c}{n} & \multicolumn{1}{c}{k} & \multicolumn{1}{c}{LTN} & \multicolumn{1}{c}{STN} & \multicolumn{1}{c}{ASP} & \multicolumn{1}{c}{ADP} & \multicolumn{1}{c}{AR} & \multicolumn{1}{c}{MDP} & \multicolumn{1}{c}{AR/MDP} & \multicolumn{1}{c}{SR} & \multicolumn{1}{c}{PnL} & \multicolumn{1}{c}{WR} & \multicolumn{1}{c}{AP/AL} \\
    \midrule
    5     & 1     & 3     & 86275 & 84782 & -13.69 & -1937.17 & -48.43\% & 233.67\% & -0.21 & -2.75 & -0.43 & 34.41\% & 1.09 \\
    5     & 3     & 3     & 77495 & 72745 & -19.74 & -2453.05 & -61.33\% & 295.71\% & -0.21 & -3.25 & -0.48 & 33.50\% & 1.02 \\
    9     & 3     & 3     & 58033 & 54595 & -17.55 & -1635.29 & -40.88\% & 197.09\% & -0.21 & -2.70 & -0.40 & 35.73\% & 1.07 \\
    14    & 3     & 3     & 47621 & 45688 & -9.75 & -752.75 & -18.82\% & 92.08\% & -0.20 & -1.19 & -0.23 & 40.28\% & 1.14 \\
    19    & 3     & 3     & 42612 & 41947 & -10.64 & -744.47 & -18.61\% & 91.47\% & -0.20 & -1.30 & -0.23 & 37.97\% & 1.25 \\
    \bottomrule
    \end{tabular}%
  \label{TAB:KDJ1}}%
\end{table}%

\begin{table}[htbp]\scalebox{0.8}{
  \centering
  \caption{Performance of stationary KDJ strategy with 30s data}
    \begin{tabular}{rrrrrrrrrrrrrr}
    \toprule
    \multicolumn{1}{c}{m} & \multicolumn{1}{c}{n} & \multicolumn{1}{c}{k} & \multicolumn{1}{c}{LTN} & \multicolumn{1}{c}{STN} & \multicolumn{1}{c}{ASP} & \multicolumn{1}{c}{ADP} & \multicolumn{1}{c}{AR} & \multicolumn{1}{c}{MDP} & \multicolumn{1}{c}{AR/MDP} & \multicolumn{1}{c}{SR} & \multicolumn{1}{c}{PnL} & \multicolumn{1}{c}{WR} & \multicolumn{1}{c}{AP/AL} \\
    \midrule
    5     & 1     & 3     & 42679 & 41162 & -11.22 & -778.11 & -19.45\% & 103.34\% & -0.19 & -1.21 & -0.23 & 36.72\% & 1.31 \\
    5     & 3     & 3     & 37831 & 35325 & -10.97 & -664.07 & -16.60\% & 87.58\% & -0.19 & -0.93 & -0.18 & 38.05\% & 1.32 \\
    9     & 3     & 3     & 27078 & 25580 & 6.85  & 298.21 & 7.46\% & 26.72\% & 0.28  & 0.48  & 0.10  & 46.82\% & 1.25 \\
    14    & 3     & 3     & 21687 & 21066 & 32.64 & 1154.24 & 28.86\% & 11.76\% & 2.45  & 1.99  & 0.35  & 50.62\% & 1.50 \\
    19    & 3     & 3     & 19450 & 19298 & 23.34 & 748.04 & 18.70\% & 17.73\% & 1.05  & 1.40  & 0.26  & 49.05\% & 1.38 \\
    \bottomrule
    \end{tabular}%
  \label{TAB:KDJ2}}%
\end{table}%

\begin{table}[htbp]\scalebox{0.8}{
  \centering
  \caption{Performance of stationary KDJ strategy with 60s data}
    \begin{tabular}{rrrrrrrrrrrrrr}
    \toprule
    \multicolumn{1}{c}{m} & \multicolumn{1}{c}{n} & \multicolumn{1}{c}{k} & \multicolumn{1}{c}{LTN} & \multicolumn{1}{c}{STN} & \multicolumn{1}{c}{ASP} & \multicolumn{1}{c}{ADP} & \multicolumn{1}{c}{AR} & \multicolumn{1}{c}{MDP} & \multicolumn{1}{c}{AR/MDP} & \multicolumn{1}{c}{SR} & \multicolumn{1}{c}{PnL} & \multicolumn{1}{c}{WR} & \multicolumn{1}{c}{AP/AL} \\
    \midrule
    5     & 1     & 3     & 20035 & 19648 & 41.31 & 1355.88 & 33.90\% & 11.68\% & 2.90  & 2.24  & 0.38  & 51.94\% & 1.49 \\
    5     & 3     & 3     & 17523 & 17376 & 17.83 & 514.79 & 12.87\% & 21.28\% & 0.60  & 0.82  & 0.16  & 51.03\% & 1.14 \\
    9     & 3     & 3     & 12293 & 12563 & 56.58 & 1163.23 & 29.08\% & 8.40\% & 3.46  & 2.10  & 0.36  & 52.03\% & 1.43 \\
    14    & 3     & 3     & 10138 & 10300 & 46.85 & 792.06 & 19.80\% & 13.61\% & 1.45  & 1.50  & 0.27  & 50.70\% & 1.33 \\
    19    & 3     & 3     & 8987  & 9503  & 43.34 & 662.88 & 16.57\% & 18.11\% & 0.92  & 1.28  & 0.25  & 49.05\% & 1.37 \\
    \bottomrule
    \end{tabular}%
  \label{TAB:KDJ3}}%
\end{table}%

\begin{table}[htbp]\scalebox{0.8}{
  \centering
  \caption{Performance of stationary Bollinger band strategy with 15s data}
    \begin{tabular}{rrrrrrrrrrrrrr}
    \toprule
    \multicolumn{1}{c}{n} & \multicolumn{1}{c}{K} & \multicolumn{1}{c}{LTN} & \multicolumn{1}{c}{STN} & \multicolumn{1}{c}{ASP} & \multicolumn{1}{c}{ADP} & \multicolumn{1}{c}{AR} & \multicolumn{1}{c}{MDP} & \multicolumn{1}{c}{AR/MDP} & \multicolumn{1}{c}{SR} & \multicolumn{1}{c}{PnL} & \multicolumn{1}{c}{WR} & \multicolumn{1}{c}{AP/AL} \\
    \midrule
    20    & 0.1   & 105419 & 105934 & -8.00  & -1398.11  & -34.95\% & 137.53\% & -0.25  & -1.36  & -0.23  & 38.79\% & 1.20  \\
    20    & 0.5   & 99120 & 100739 & -7.22  & -1193.45  & -29.84\% & 125.44\% & -0.24  & -1.33  & -0.23  & 36.89\% & 1.32  \\
    20    & 1     & 81557 & 83427 & -7.70  & -1051.07  & -26.28\% & 120.84\% & -0.22  & -1.32  & -0.23  & 35.15\% & 1.41  \\
    20    & 1.5   & 58513 & 60634 & -10.23  & -1008.09  & -25.20\% & 121.53\% & -0.21  & -1.53  & -0.27  & 33.58\% & 1.44  \\
    20    & 2     & 37774 & 39203 & -15.57  & -991.46  & -24.79\% & 119.82\% & -0.21  & -1.89  & -0.32  & 33.50\% & 1.33  \\
    20    & 2.5   & 22171 & 22969 & -22.40  & -836.48  & -20.91\% & 101.13\% & -0.21  & -2.32  & -0.37  & 32.34\% & 1.31  \\
    30    & 0.1   & 86400 & 86326 & -5.19  & -742.03  & -18.55\% & 95.12\% & -0.20  & -0.77  & -0.14  & 40.28\% & 1.27  \\
    30    & 0.5   & 80918 & 82258 & -6.87  & -926.85  & -23.17\% & 107.11\% & -0.22  & -1.10  & -0.19  & 38.21\% & 1.30  \\
    30    & 1     & 67055 & 68568 & -6.24  & -699.95  & -17.50\% & 89.12\% & -0.20  & -0.87  & -0.17  & 35.81\% & 1.48  \\
    30    & 1.5   & 49164 & 50905 & -8.88  & -734.89  & -18.37\% & 96.46\% & -0.19  & -1.06  & -0.20  & 33.91\% & 1.56  \\
    30    & 2     & 32324 & 33681 & -14.73  & -804.32  & -20.11\% & 97.34\% & -0.21  & -1.59  & -0.27  & 34.66\% & 1.36  \\
    30    & 2.5   & 19341 & 20008 & -17.96  & -584.62  & -14.62\% & 71.01\% & -0.21  & -1.62  & -0.28  & 33.83\% & 1.41  \\
    60    & 0.1   & 60853 & 60812 & 0.74  & 74.24  & 1.86\% & 53.00\% & 0.04  & 0.08  & 0.02  & 42.51\% & 1.36  \\
    60    & 0.5   & 58017 & 58901 & -1.66  & -160.40  & -4.01\% & 59.58\% & -0.07  & -0.19  & -0.04  & 40.36\% & 1.42  \\
    60    & 1     & 49326 & 50797 & -10.55  & -873.70  & -21.84\% & 105.47\% & -0.21  & -1.22  & -0.21  & 37.39\% & 1.31  \\
    60    & 1.5   & 36502 & 37864 & -18.09  & -1113.00  & -27.83\% & 133.42\% & -0.21  & -1.74  & -0.30  & 34.99\% & 1.29  \\
    60    & 2     & 24291 & 25144 & -23.67  & -967.89  & -24.20\% & 116.98\% & -0.21  & -1.87  & -0.32  & 32.01\% & 1.44  \\
    60    & 2.5   & 15041 & 15184 & -29.31  & -732.66  & -18.32\% & 88.58\% & -0.21  & -1.92  & -0.32  & 34.00\% & 1.30  \\
    120   & 0.1   & 42600 & 42783 & 20.77  & 1467.15  & 36.68\% & 22.01\% & 1.67  & 1.45  & 0.26  & 48.55\% & 1.44  \\
    120   & 0.5   & 42807 & 43081 & 9.35  & 664.12  & 16.60\% & 33.80\% & 0.49  & 0.70  & 0.14  & 42.76\% & 1.55  \\
    120   & 1     & 37480 & 38854 & -12.16  & -767.74  & -19.19\% & 93.15\% & -0.21  & -1.03  & -0.19  & 37.14\% & 1.37  \\
    120   & 1.5   & 28432 & 29626 & -26.50  & -1272.66  & -31.82\% & 152.53\% & -0.21  & -2.08  & -0.34  & 34.99\% & 1.22  \\
    120   & 2     & 19059 & 19758 & -38.71  & -1242.78  & -31.07\% & 149.87\% & -0.21  & -2.50  & -0.39  & 33.09\% & 1.22  \\
    120   & 2.5   & 12060 & 12107 & -38.08  & -761.24  & -19.03\% & 92.16\% & -0.21  & -1.92  & -0.33  & 34.41\% & 1.26  \\
    \bottomrule
    \end{tabular}%
  \label{TAB:Boll1}}%
\end{table}%

\begin{table}[htbp]\scalebox{0.8}{
  \centering
  \caption{Performance of stationary Bollinger band strategy with 30s data}
    \begin{tabular}{rrrrrrrrrrrrrr}
    \toprule
    \multicolumn{1}{c}{n} & \multicolumn{1}{c}{K} & \multicolumn{1}{c}{LTN} & \multicolumn{1}{c}{STN} & \multicolumn{1}{c}{ASP} & \multicolumn{1}{c}{ADP} & \multicolumn{1}{c}{AR} & \multicolumn{1}{c}{MDP} & \multicolumn{1}{c}{AR/MDP} & \multicolumn{1}{c}{SR} & \multicolumn{1}{c}{PnL} & \multicolumn{1}{c}{WR} & \multicolumn{1}{c}{AP/AL} \\
    \midrule
    20    & 0.1   & 51228 & 51457 & 6.48  & 550.67  & 13.77\% & 34.20\% & 0.40  & 0.60  & 0.12  & 44.83\% & 1.38  \\
    20    & 0.5   & 48725 & 49425 & -1.19  & -96.43  & -2.41\% & 50.16\% & -0.05  & -0.12  & -0.02  & 42.27\% & 1.32  \\
    20    & 1     & 40292 & 41214 & -3.50  & -235.93  & -5.90\% & 55.06\% & -0.11  & -0.32  & -0.06  & 40.45\% & 1.37  \\
    20    & 1.5   & 29420 & 30293 & -7.28  & -359.55  & -8.99\% & 60.82\% & -0.15  & -0.60  & -0.12  & 39.04\% & 1.36  \\
    20    & 2     & 19101 & 19388 & -10.12  & -322.33  & -8.06\% & 45.90\% & -0.18  & -0.71  & -0.14  & 37.80\% & 1.40  \\
    20    & 2.5   & 10918 & 11222 & -29.59  & -541.94  & -13.55\% & 67.90\% & -0.20  & -1.66  & -0.29  & 35.81\% & 1.25  \\
    30    & 0.1   & 42046 & 42225 & 9.40  & 655.04  & 16.38\% & 27.29\% & 0.60  & 0.73  & 0.14  & 44.25\% & 1.46  \\
    30    & 0.5   & 40108 & 40511 & 4.45  & 297.02  & 7.43\% & 34.22\% & 0.22  & 0.36  & 0.07  & 43.01\% & 1.43  \\
    30    & 1     & 33694 & 34657 & -1.56  & -87.94  & -2.20\% & 41.32\% & -0.05  & -0.13  & -0.03  & 40.94\% & 1.40  \\
    30    & 1.5   & 24891 & 25847 & -10.72  & -449.93  & -11.25\% & 57.02\% & -0.20  & -0.72  & -0.14  & 38.54\% & 1.36  \\
    30    & 2     & 16346 & 16746 & -16.68  & -456.67  & -11.42\% & 55.63\% & -0.21  & -0.96  & -0.18  & 35.81\% & 1.45  \\
    30    & 2.5   & 9780  & 9822  & -26.06  & -422.58  & -10.56\% & 51.09\% & -0.21  & -1.14  & -0.22  & 36.81\% & 1.31  \\
    60    & 0.1   & 29747 & 29702 & 29.84  & 1467.25  & 36.68\% & 12.04\% & 3.05  & 1.48  & 0.27  & 47.97\% & 1.48  \\
    60    & 0.5   & 29495 & 29789 & 23.45  & 1150.12  & 28.75\% & 17.66\% & 1.63  & 1.24  & 0.24  & 44.58\% & 1.63  \\
    60    & 1     & 25672 & 26567 & -2.02  & -87.30  & -2.18\% & 48.11\% & -0.05  & -0.12  & -0.02  & 41.52\% & 1.37  \\
    60    & 1.5   & 19533 & 20344 & -28.02  & -924.07  & -23.10\% & 111.38\% & -0.21  & -1.55  & -0.27  & 37.39\% & 1.22  \\
    60    & 2     & 13032 & 13346 & -39.03  & -851.61  & -21.29\% & 102.93\% & -0.21  & -1.78  & -0.31  & 35.24\% & 1.26  \\
    60    & 2.5   & 8054  & 7969  & -28.01  & -371.22  & -9.28\% & 45.90\% & -0.20  & -0.99  & -0.20  & 36.48\% & 1.39  \\
    120   & 0.1   & 22234 & 22325 & 64.37  & 2372.41  & 59.31\% & 6.64\% & 8.93  & 2.33  & 0.40  & 50.79\% & 1.62  \\
    120   & 0.5   & 22333 & 22916 & 40.19  & 1504.27  & 37.61\% & 9.69\% & 3.88  & 1.63  & 0.30  & 46.57\% & 1.61  \\
    120   & 1     & 20223 & 21423 & 6.06  & 208.83  & 5.22\% & 37.81\% & 0.14  & 0.26  & 0.05  & 42.43\% & 1.42  \\
    120   & 1.5   & 15788 & 17078 & -14.03  & -381.29  & -9.53\% & 50.79\% & -0.19  & -0.57  & -0.11  & 38.79\% & 1.39  \\
    120   & 2     & 11161 & 11440 & -21.21  & -396.48  & -9.91\% & 50.44\% & -0.20  & -0.77  & -0.15  & 37.47\% & 1.41  \\
    120   & 2.5   & 7228  & 7364  & -16.06  & -193.80  & -4.84\% & 26.69\% & -0.18  & -0.49  & -0.10  & 37.22\% & 1.49  \\
    \bottomrule
    \end{tabular}%
  \label{TAB:Boll2}}%
\end{table}%

\begin{table}[htbp]\scalebox{0.8}{
  \centering
  \caption{Performance of stationary Bollinger band strategy with 60s data}
    \begin{tabular}{rrrrrrrrrrrrrr}
    \toprule
    \multicolumn{1}{c}{n} & \multicolumn{1}{c}{K} & \multicolumn{1}{c}{LTN} & \multicolumn{1}{c}{STN} & \multicolumn{1}{c}{ASP} & \multicolumn{1}{c}{ADP} & \multicolumn{1}{c}{AR} & \multicolumn{1}{c}{MDP} & \multicolumn{1}{c}{AR/MDP} & \multicolumn{1}{c}{SR} & \multicolumn{1}{c}{PnL} & \multicolumn{1}{c}{WR} & \multicolumn{1}{c}{AP/AL} \\
    \midrule
    20    & 0.1   & 25553 & 25588 & 14.26  & 603.33  & 15.08\% & 25.98\% & 0.58  & 0.64  & 0.13  & 46.48\% & 1.30  \\
    20    & 0.5   & 24336 & 24745 & 6.63  & 268.98  & 6.72\% & 30.35\% & 0.22  & 0.33  & 0.07  & 43.84\% & 1.37  \\
    20    & 1     & 20693 & 21070 & -6.14  & -212.26  & -5.31\% & 44.03\% & -0.12  & -0.31  & -0.06  & 42.76\% & 1.25  \\
    20    & 1.5   & 15150 & 15395 & -11.31  & -285.66  & -7.14\% & 42.79\% & -0.17  & -0.52  & -0.10  & 41.03\% & 1.28  \\
    20    & 2     & 9630  & 9799  & -23.85  & -383.28  & -9.58\% & 54.03\% & -0.18  & -0.90  & -0.17  & 39.95\% & 1.23  \\
    20    & 2.5   & 5539  & 5484  & -3.84  & -34.99  & -0.87\% & 28.53\% & -0.03  & -0.10  & -0.02  & 41.60\% & 1.36  \\
    30    & 0.1   & 20998 & 20990 & 19.75  & 686.00  & 17.15\% & 23.08\% & 0.74  & 0.73  & 0.14  & 47.06\% & 1.30  \\
    30    & 0.5   & 20526 & 20655 & 25.96  & 884.32  & 22.11\% & 21.10\% & 1.05  & 1.02  & 0.20  & 46.15\% & 1.45  \\
    30    & 1     & 17609 & 18201 & -1.42  & -41.99  & -1.05\% & 29.16\% & -0.04  & -0.06  & -0.01  & 42.93\% & 1.30  \\
    30    & 1.5   & 13204 & 13556 & -23.31  & -515.98  & -12.90\% & 65.01\% & -0.20  & -0.91  & -0.17  & 40.45\% & 1.22  \\
    30    & 2     & 8695  & 8830  & -36.91  & -535.04  & -13.38\% & 69.82\% & -0.19  & -1.21  & -0.23  & 39.21\% & 1.19  \\
    30    & 2.5   & 5167  & 5040  & -5.80  & -48.98  & -1.22\% & 26.08\% & -0.05  & -0.14  & -0.03  & 39.54\% & 1.46  \\
    60    & 0.1   & 15543 & 15653 & 69.07  & 1782.13  & 44.55\% & 9.38\% & 4.75  & 1.79  & 0.32  & 49.96\% & 1.48  \\
    60    & 0.5   & 15520 & 15913 & 44.54  & 1157.97  & 28.95\% & 10.79\% & 2.68  & 1.31  & 0.24  & 46.82\% & 1.48  \\
    60    & 1     & 13853 & 14703 & 13.14  & 310.32  & 7.76\% & 16.77\% & 0.46  & 0.41  & 0.08  & 43.34\% & 1.41  \\
    60    & 1.5   & 10790 & 11455 & -4.62  & -85.06  & -2.13\% & 25.61\% & -0.08  & -0.13  & -0.03  & 41.36\% & 1.37  \\
    60    & 2     & 7557  & 7743  & -15.19  & -192.26  & -4.81\% & 36.96\% & -0.13  & -0.38  & -0.08  & 38.38\% & 1.46  \\
    60    & 2.5   & 4721  & 4764  & 8.27  & 64.91  & 1.62\% & 19.12\% & 0.08  & 0.17  & 0.04  & 39.21\% & 1.59  \\
    120   & 0.1   & 13024 & 13140 & 50.20  & 1086.40  & 27.16\% & 10.04\% & 2.71  & 1.13  & 0.21  & 47.97\% & 1.36  \\
    120   & 0.5   & 12818 & 13562 & 46.51  & 1014.84  & 25.37\% & 12.98\% & 1.95  & 1.14  & 0.21  & 45.82\% & 1.48  \\
    120   & 1     & 11888 & 13011 & 10.97  & 225.86  & 5.65\% & 18.28\% & 0.31  & 0.28  & 0.06  & 43.09\% & 1.39  \\
    120   & 1.5   & 9733  & 10497 & 9.93  & 166.20  & 4.16\% & 16.73\% & 0.25  & 0.25  & 0.05  & 42.02\% & 1.44  \\
    120   & 2     & 6920  & 7477  & -0.80  & -9.58  & -0.24\% & 20.36\% & -0.01  & -0.02  & 0.00  & 39.87\% & 1.49  \\
    120   & 2.5   & 4551  & 4712  & 41.09  & 314.79  & 7.87\% & 14.53\% & 0.54  & 0.78  & 0.16  & 39.87\% & 1.75  \\
    \bottomrule
    \end{tabular}%
  \label{TAB:Boll3}}%
\end{table}%

\end{appendices}

\end{document}